\documentclass[twocolumn]{aastex631}

\accepted{December 19, 2023}

\submitjournal{APJ}

\graphicspath{{./}{figures/}}
\usepackage{subfigure}
\usepackage{amsmath}
\usepackage{multirow}
\usepackage{comment}
\usepackage{dirtytalk}

\begin{document}

\title{\Large Deep Chandra observations of Abell 2495: a possible sloshing-regulated feedback cycle in a triple-offset galaxy cluster}

\author[0000-0002-0327-5929]{L. Rosignoli}
\affiliation{Dipartimento di Fisica e Astronomia (DIFA), Università di Bologna, via Gobetti 93/2, I-40129 Bologna, Italy}
\affiliation{Istituto Nazionale di Astrofisica (INAF), Osservatorio di Astrofisica e Scienza dello Spazio, via P. Gobetti 93/3, I-40129 Bologna, Italy}

\author[0000-0001-5338-4472]{F. Ubertosi}
\affiliation{Dipartimento di Fisica e Astronomia (DIFA), Università di Bologna, via Gobetti 93/2, I-40129 Bologna, Italy}
\affiliation{Istituto Nazionale di Astrofisica (INAF), Osservatorio di Astrofisica e Scienza dello Spazio, via P. Gobetti 93/3, I-40129 Bologna, Italy}

%\collaboration{6}{(AAS Journals Data Editors)}

\author[0000-0002-0843-3009]{M. Gitti}
\affiliation{Dipartimento di Fisica e Astronomia (DIFA), Università di Bologna, via Gobetti 93/2, I-40129 Bologna, Italy}
\affiliation{Istituto Nazionale di Astrofisica (INAF) - Istituto di Radioastronomia (IRA), via Gobetti 101, I-40129 Bologna, Italy}

\author[0000-0001-9807-8479]{F. Brighenti}
%\altaffiliation{AASTeX v6+ programmer}
\affiliation{Dipartimento di Fisica e Astronomia (DIFA), Università di Bologna, via Gobetti 93/2, I-40129 Bologna, Italy}
\affiliation{University of California/Lick Observatory, Departement of Astronomy and Astrophysics, University of California, Santa Cruz, CA 95064, USA}

\author[0000-0002-9711-5554]{T. Pasini}
\affiliation{Istituto Nazionale di Astrofisica (INAF) - Istituto di Radioastronomia (IRA), via Gobetti 101, I-40129 Bologna, Italy}

\author[0000-0002-5671-6900]{E. O'Sullivan}
\affiliation{Center for Astrophysics $|$ Harvard $\&$ Smithsonian, 60 Garden Street, Cambridge, MA 02138, USA}

\author[0000-0002-9112-0184]{F. Gastaldello}
\affiliation{INAF-IASF Milano, via E. Bassini 15, I-20133 Milano, Italy}

\author[0000-0003-2754-9258]{M. Gaspari}
\affiliation{Dept. of Astrophysical Sciences, Princeton University, 4 Ivy Lane, Princeton, NJ 08544, USA.}

\author[0000-0002-8341-342X]{P. Temi}
\affiliation{Astrophysics Branch, NASA-Ames Research Center, MS 245-6, Moffett Field, CA 94035, USA}



\begin{abstract}

We present the analysis of new, deep $Chandra$ observations (130~ks) of the galaxy cluster Abell~2495. This object is known for the presence of a triple offset between the peaks of the intracluster medium (ICM), the brightest cluster galaxy (BCG), and the warm gas glowing in H$\alpha$ line. The new $Chandra$ data confirm that the X-ray emission peak is located at a distance of $\sim$6.2 kpc from the BCG, and at $\sim$3.9 kpc from the H$\alpha$ emission peak. Moreover, we identify two generations of X-ray cavities in the ICM, likely inflated by the central radio galaxy activity. Through a detailed morphological and spectral analysis we determine that the power of the AGN outbursts ($P_{cav} = 4.7\pm1.3\times10^{43}$~erg~s$^{-1}$) is enough to counterbalance the radiative losses from ICM cooling ($L_{cool} = 5.7\pm0.1\times10^{43}$~erg~s$^{-1}$). This indicates that, despite a fragmented cooling core, Abell~2495 still harbors an effective feedback cycle. We argue that the offsets are most likely caused by sloshing of the ICM, supported by the presence of spiral structures and a probable cold front in the gas at $\sim$58 kpc east of the center. Ultimately, we find that the outburst interval between the two generations of X-ray cavities is of the order of the dynamical sloshing timescale, as already hinted from the previous $Chandra$ snapshot. We thus speculate that sloshing may be able to regulate the timescales of AGN feedback in Abell~2495, by periodically fuelling the central AGN.

\end{abstract}



\section{Introduction} \label{sec:intro}
Galaxy clusters are multi-component systems with a Dark Matter (DM) halo, member galaxies, and the intracluster medium (ICM). As the hot ionized ICM radiates away its energy through thermal X-ray emission, it will contract to maintain its dynamical equilibrium, setting up a pressure driven inflow also known as a cooling flow \emph{cooling flow process} \citep{fabian_94a}. 
In the “central galaxy paradigm” \citep{bosch_05,cui_16} the brightest cluster galaxy (BCG) is expected to be located at the center of the host cluster, as well as the peak of X-ray emission from the hot ICM and the emission lines from cooling gas (e.g., H$\alpha$ from the warm ionized phase or rotational CO lines from molecular gas), all located at the bottom of the cluster potential well. In this situation the cooling gas could feed the super massive black hole (SMBH) hosted in the BCG, triggering its feedback. Such activity can heat up again the surrounding gas with many processes (e.g. shocks and cavities) and a tight trade off between cooling and heating is created, usually dubbed the AGN feedback cycle \citep{mcnamara_07,mcnamara_12}.\\
However, in case of interactions with other clusters or groups, or after a powerful active galactic nucleus (AGN) outburst, all these components are likely to shift, leading to spatial offsets between them. Recent studies confirm that the relative position between the BCG, the X-ray core and the line emission peaks can be influenced by the dynamic state of the cluster.\\
\citet{sanderson_09} studied 65 X-ray selected clusters from the Local Cluster Substructure Survey (LoCuSS), finding a clear correlation among the projected offset between the X-ray centroid and the BCG and the logarithmic slope of the intracluster medium (ICM) density profile ($\alpha$) at 0.04 r$_{500}$, implying that more dynamically disturbed clusters have weaker cool cores. Moreover, \citet{hudson_10} studied 64 clusters belonging to the HIghest X-ray FLUx Galaxy Cluster Sample (HIFLUGCS) with X-ray data from $Chandra$ observations, finding that a large projected separation between the BCG and the X-ray peak is a good indicator of a major merger. 
\\The relative positions of the BCG, the X-ray peak and the line emission may also influence the thermodynamics of the gas at the cluster center. \citet{hamer_12} studied three clusters (Abell 1991, Abell 3444 and Ophiuchus), in which offsets between the BCG and the cooling peak are present, in order to investigate the connection between the cooling of the ICM, the cold gas being deposited and the central galaxy. In A1991, the detection of CO(2–1) emission both on the BCG and on the peak of the soft X-ray emission suggested that cooling still occurs in the core despite being offset from the BCG. Even though these occurrences seem to be rare \citep[e.g. 4/73 in the sample of][]{hamer_16}, they provide a unique opportunity to both directly constrain the process of gas cooling far from the BCG, and to study the sustainability of the AGN feedback in such environments. An example is the work of \citet{pasini_21} focused on the AGN feedback cycle of Abell 1668, a cluster with a significant BCG -- X-ray peak offset. They detected two putative X-ray cavities, and they evaluated the position of the cluster in the cooling luminosity-cavity power parameter space, finding that the AGN energy injection is able to prevent cooling, and it is likely that the offset did not break the AGN feedback cycle. This situation is also called as \emph{self-regulated duty cycle scenario} \citep[e.g.][for a review.]{gaspari_20}

In this context, the cluster under investigation in this study, Abell 2495 (hereafter A2495, RA:22 50 17.10; DEC:+10 55 01.20), is a relevant system. It has been selected from the ROSAT Brightest Cluster Sample \citep[BCS][]{ebeling_98} as a cluster with high X-ray flux ($F_X = 1.18\times10^{-11}$ erg cm$^{-2}$ s$^{-1}$) and H$\alpha$ luminosity \citep{crawford_99} $L_{H\alpha} > 10^{40}$ erg s$^{-1}$ \citep[for a similar selection criteria see][]{ettori_13,pasini_19,pasini_21,ubertosi_23}. \citet{pasini_19} performed a multi-frequency study of this cluster using a 8 ks $Chandra$ snapshot, 1.4 GHz and 5 GHz EVLA observations, optical images from the Hubble Space Telescope (HST) archive, and H$\alpha$ data from \citet{hamer_16}. They characterized the radio activity of the BCG, finding that it is an FR-I radio galaxy with 1.4 GHz luminosity of $2.18 \times 10^{23}$ W Hz$^{-1}$ and size of $\sim$14 kpc. The X-ray analysis of the ICM suggested that this cluster has a cool core, where a significant spatial offset between the X-ray peak and the BCG ($\sim 6$ kpc) is present. An H$\alpha$ nebula encasing a dust filament connects the peak of the X-ray emission to the peak of the BCG radio-optical continuum, where the dust peak is located. Interestingly, the X-ray emission peak is also slightly offset ($\sim$ 3.5 kpc) from that of the H$\alpha$ emission, suggesting that a merger or some other strong disturbance has separated the BCG nucleus from both the hot gas and the nebular emission. Since such spatial offsets are likely to occur in many clusters at some point in their evolution, it is important to investigate whether they can affect the feedback cycle. The snapshot Chandra images hint at the presence of cavities inflated by the recurrent AGN activity. The analysis of these putative cavities suggested that the cavity power $P_{cav}$ can match the ICM radiative losses. This possibly indicated that AGN feedback can still balance the cooling process even in the presence of a spatial offset between the bulk of the cooling gas and the supermassive black hole (SMBH) at the center of the BCG \citep{pasini_19}. However, these preliminary arguments need further confirmation from deeper data. This cluster is thus a perfect case to investigate the resilience of the AGN feedback cycle even in a disturbed and dynamic environment. 

In this work we investigate in detail the physical and dynamical state of the ICM at the center of A2495 by exploiting the deep, 130 ks $Chandra$ observations that we obtained in Cycle 22 (see Table \ref{tab:dat}). We describe the data set and the reduction process in Section \ref{sec:data}. The results of the morphological and spectral analysis are reported respectively in Section \ref{sec:an-morph} and Section \ref{sec:an-spec}. We discuss the results in Section \ref{sec:disc} and we briefly summarize our conclusion in Section \ref{sec:concl}. The cosmological parameters assumed for this work are: $H_0=70$ km s$^{-1}$ Mpc$^{-1}$, $\Omega_0=0.3$, $\Omega_{\Lambda}=0.7$. The redshift of the cluster is $z=0.07923$, that implies a scale of $1.5$ kpc arcsec$^{-1}$.

\section{The Data} \label{sec:data}

\begin{figure*}
    \centering
    \subfigure[]{
    \includegraphics[scale=0.27]{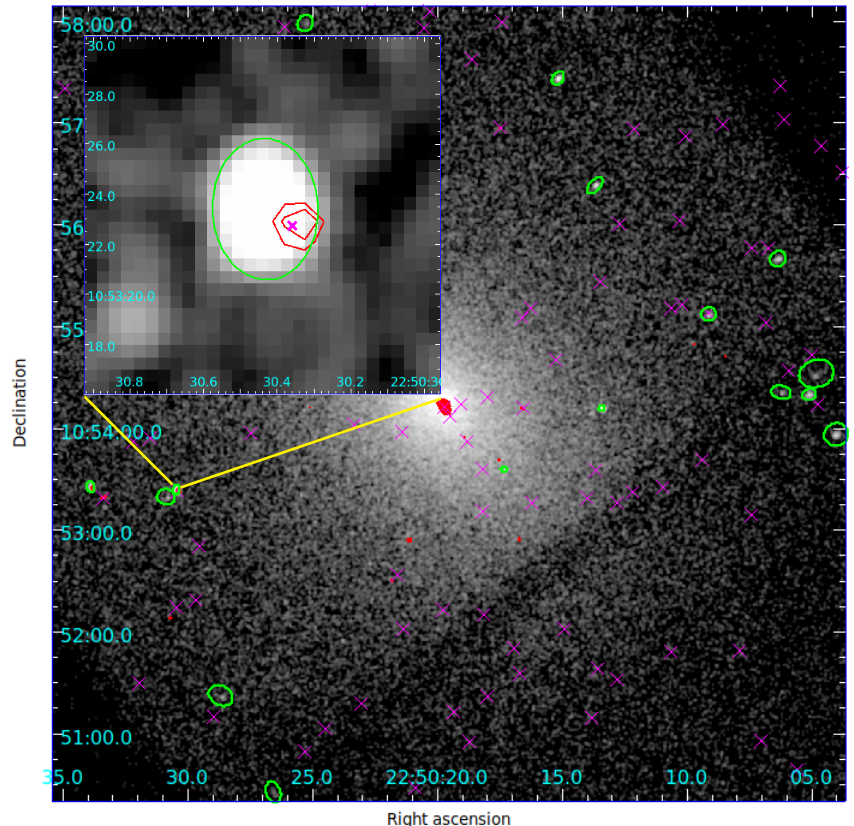}}
    \subfigure[]{
    \includegraphics[scale=0.27]{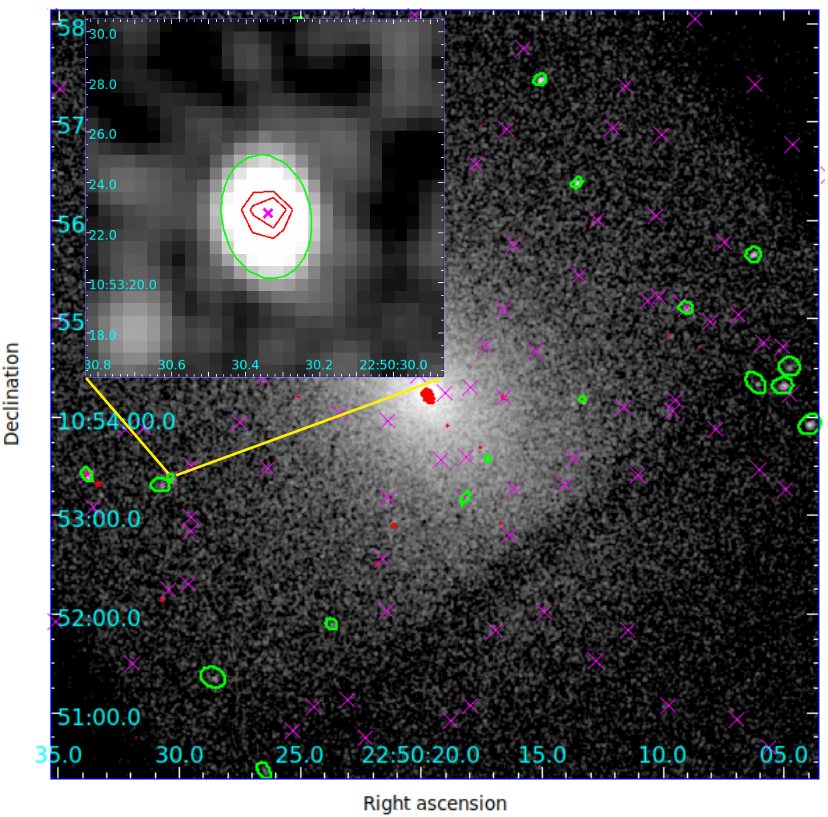}}
    \caption{(a) Final 0.5-7 keV mosaic from the standard astrometric correction procedure (using the USNO-A2.0 catalogue). (b) Final 0.5-7 keV mosaic after the new astrometric correction procedure (using the GAIA-DR2 catalogue). In both images the regions of the point sources found by \texttt{wavdetect} are overlaid in green, the radio contours at 1.4 GHz in red (1.29"$\times$1.12", at -3,3,6,12,24,48 $\times$ rms, with an rms noise of $10\,\mu$Jy beam$^{-1}$, \citet{pasini_19}), and the catalog coordinates with the magenta crosses. The comparison between the two panels shows the improvement in astrometric accuracy by adopting the second method, rather than the standard one (see Section \ref{sec:data} for details).}
    \label{fig:offset}
\end{figure*}

The X-ray data for A2495 consist of the previous observation acquired during the 12th cycle (ObsID: 12876), and six new \emph{Chandra} observations, acquired during the 22nd cycle (ObsID: 23849, 24277, 24278, 24279, 24650, 24659), achieving a total exposure of $\sim$ 140 ks (see Table \ref{tab:dat}). 
\begin{table}
    \centering
    \begin{tabular}{l|c|c|l|c}
    \hline
     Information & Proposal & Date & ObsId & Exp [ks] \\
    \hline
    P.I. M. Gitti & \multirow{6}*{22800391} & \multirow{6}*{09-2020} & 23849 & 33.0\\
    \cline{4-5}
    Target: A2495 & & & 24277 & 22.0 \\
    \cline{4-5}
    RA: 22 50 17.10& & & 24278 & 27.0 \\
    \cline{4-5}
    DEC: +10 55 01.20 & & & 24279 & 16.5 \\
    \cline{4-5}
    Sensor: ACIS-S & & & 24650 & 22.0\\
    \cline{4-5}
    Mode: V-FAINT & & & 24659 & 9.5 \\
    \cline{2-5}
    Tot. Exp: 138 ks & 12800143 & 12-2010 & 12876 & 8.0\\
    \hline
    \end{tabular}
\caption{Summary of the $Chandra$ data analyzed in this work.}
\label{tab:dat}
\end{table}
All the observations have been acquired using the ACIS-S sensor in VFAINT mode. The calibration has been performed with \texttt{CIAO-4.12} and \texttt{CALDB-4.9.3}. First the \texttt{chandra$\_$repro} script performed the standard processing steps recommended by \emph{Chandra X-ray Observatory} (CXC) producing the bad pixel file, the level-2 event file and the appropriate response files. To achieve a high astrometric accuracy, which is essential in order to study the central offsets detected in this cluster, we performed an astrometric correction of the Chandra data (with the aim of improving the nominal pointing accuracy of 0.4’’). We start by following the standard CXC's threads\footnote{\url{https://cxc.cfa.harvard.edu/ciao/threads/reproject_aspect/}}, cross-matching the point sources (detected using the \texttt{wavdetect} tool) in the longest observation (OsbID 23849) with the external catalog USNO-A2.0, and then aligning the other ObsIDs with the corrected one. With this method, however, a significant and non physical offset ($\sim0.5"$) between the X-ray and radio data was introduced (Figure \ref{fig:offset}a). Therefore, we tried a different approach: first we aligned each observation with the longest one, then we stacked them to create a mosaic. This file (rather than the longest ObsID alone) was used to extract the positions of the point sources within the field of view. We matched the coordinates of the detected point sources with the external catalog GAIA-DR2 \citep{gaia_18}, that has an astrometric accuracy of $\Delta\theta\approx$ 0.04 mas \citep{gaia_18}, compared with the $\Delta\theta\approx$ 250 mas of the USNO-A2.0 catalogue. With this procedure, the X-ray point source are now coincident with the reference catalogue as shown in Figure \ref{fig:offset}b.\\
\noindent Then, the time intervals containing background flares have been cleaned up exploiting the \texttt{deflare} task. The net exposure after this process is $\sim$ 120 ks. We used the \texttt{blanksky} task to select and reproject the background data for each observation of the data set and chip. These files were normalized to match the 9-12 keV count rate of the observation. These calibration steps were done for each observation and for both the S3 and S2 chips, as the cluster falls within both.

\section{Morphological Analysis} \label{sec:an-morph}
\begin{figure}
    \centering
    \includegraphics[width=\linewidth]{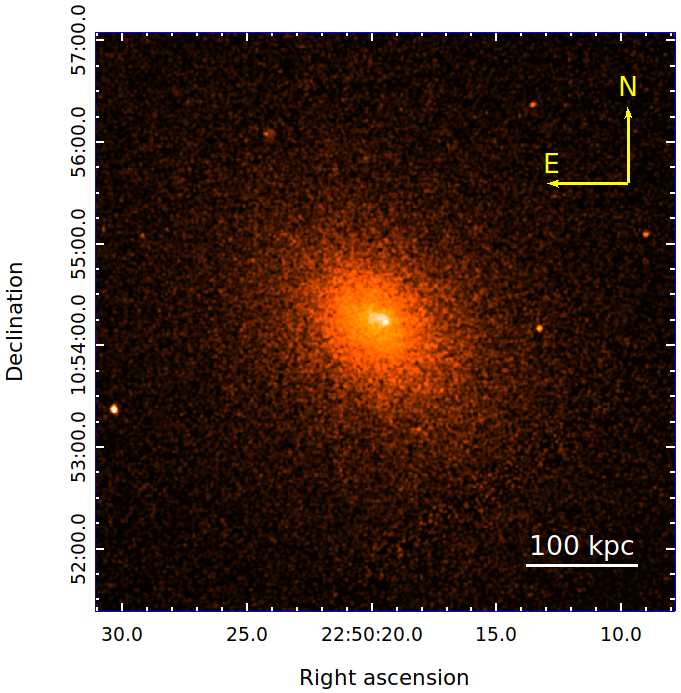}
    \caption{0.5-2 keV mosaiced $Chandra$ image, smoothed with a gaussian kernel of $\sigma=3$".}
    \label{fig:A2495-global}
\end{figure}

In order to study the global ICM emission, we used the \texttt{merge$\_$obs} command to stack all the registered images together, and then we executed \texttt{fluximage} again. These steps allowed us to produce a mosaiced, exposure-corrected, 0.5-2 keV image of the cluster (see Figure \ref{fig:A2495-global}).  On large scales (hundreds of kpc), A2495 shows a bright core and no hints of recent major merger, since no other extended substructure seems to be present within the field of view.
Using the \texttt{PROFFIT} tool \texttt{ellipticity} we measured an ellipticity of $e=1-\frac{b}{a}=0.25$ (where $b$ and $a$ are the minor and major axes), and a position angle of 120$^{\circ}$ counter-clockwise from the DEC axis.

\subsection{The offsets between the X-ray peak, the BCG, and the H$\alpha$ peak}\label{sec:offsets}

\begin{figure}
    \centering
    \includegraphics[scale=0.45]{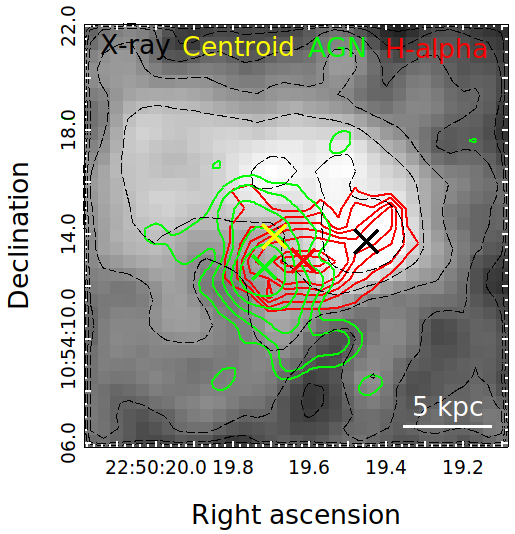}
    \caption{0.5-2 keV \emph{Chandra} image of the central region of A2495.  The radio contours at 5.0 GHz from \citet{pasini_19} are shown in green (beam of $1.1" \times 1.1"$; with an rms of $4\mu$Jy beam$^{-1}$; contours at -3,3,6,12,24,48 $\times$ rms) while the H$\alpha$ contours from \citet{hamer_16}  are shown in red 
    ($F_{H\alpha} \simeq 9.03 \pm 0.81 \cdot 10^{-16}$ erg s$^{-1}$ cm$^{-2}$; seeing of 0.95"). The black, red and green crosses represent respectively the X-ray peak, the H$\alpha$-peak and the center of the BCG. The yellow cross represents the X-ray emission centroid.}
    \label{fig:A2495-offset}
\end{figure}

Figure \ref{fig:A2495-offset} shows the very central region ($\lesssim 30$ kpc) of A2495. In order to evaluate the relative positions between the BCG and multi-phase ICM, we overlaid both the radio contour at 5 GHz \citep{pasini_19} and the H$\alpha$ contour \citep{hamer_16} on the X-ray 0.5-2 keV image. 
Two significant offsets are visible: one between the X-ray peak and the center of the BCG ($6.2\pm0.8$ kpc) and the other between the X-ray peak and the H$\alpha$-peak ($3.9\pm2.4$ kpc); see Table \ref{tab:centers}. Thus, the longer exposure confirms the previous values found by \citet{pasini_19}. 
For completeness, we also determined the position of the ICM emission centroid exploiting the \texttt{statistic} tool of \texttt{DS9}. 
%\begin{center}
    \begin{table*}
        \centering
        \begin{tabular}{|c|c|c|c|}
        \hline
            Center & $\alpha$ & $\delta$ & Offset arcsec(kpc)\\
            \hline
             X-ray peak & $\text{22:50:19.5}\pm0.5$ & $\text{10:54:13.7}\pm0.5$ & -- \\
            \hline
             BCG center & $\text{22:50:19.7}\pm0.1$ & $\text{10:54:12.7}\pm0.1$ & $4.4\pm0.5(6.2\pm0.8)$\\
             \hline
             H$\alpha$ peak & $\text{22:50:19.6}\pm1.5$ & $\text{10:54:13.0}\pm1.5$ & $2.6\pm1.6(3.9\pm2.4)$\\
        \hline
        \end{tabular}
        \caption{Coordinates of the X-ray peak (measured from the $Chandra$ data), the BCG and the H$\alpha$-peak (from \citet{pasini_19}). The last column reports the distance from the X-ray peak.}
        \label{tab:centers}
    \end{table*}
%\end{center}

\subsection{Surface Brightness Profile}\label{sec:SBP}
\begin{figure*}
    \centering
    \subfigure[]{
    \includegraphics[scale=0.36]{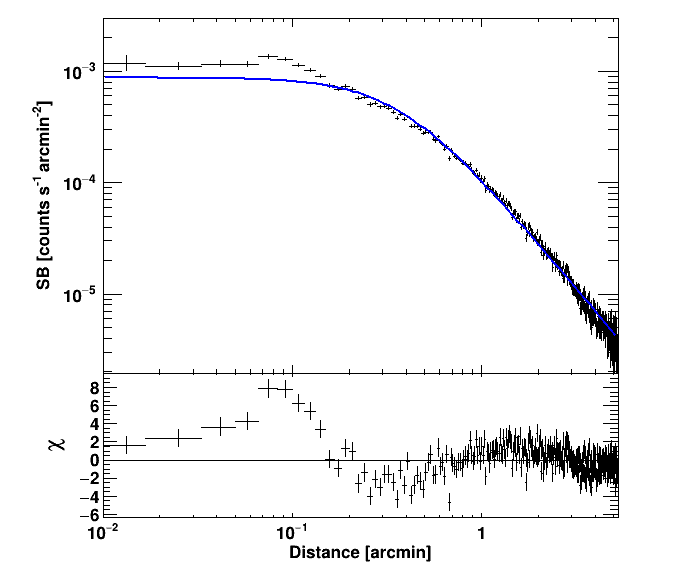}
    }
    \subfigure[]{
    \includegraphics[scale=0.36]{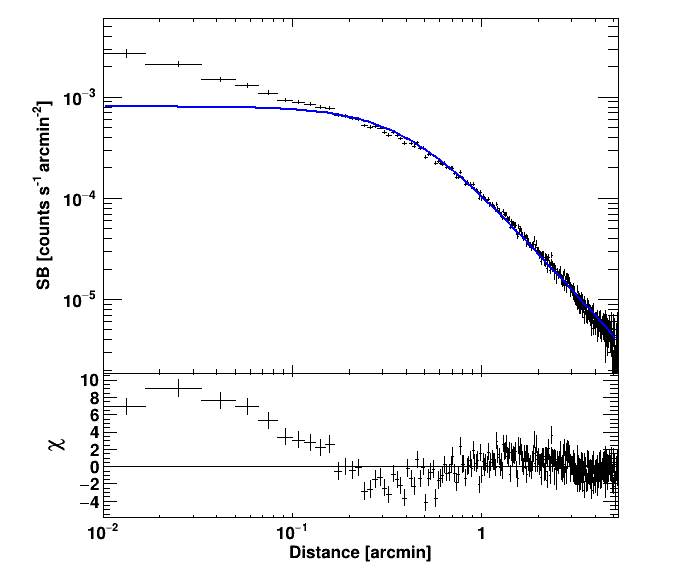}
    }
    \caption{Surface brightness radial profiles of A2495 fitted with a single $\beta$-model (blue line): (a) BCG centered; (b) X-ray peak centered. The residuals ($\frac{|\text{data}-\text{model}|}{\sigma}$) are shown in the bottom panel of each plot.}
    \label{fig:A2495-SBprofBeta}
\end{figure*}

We used \texttt{PROFFIT (v.1.5)} to extract and fit the surface brightness radial profile in series of 2"-width concentric elliptical annuli from the exposure-corrected, background-subtracted image. The selection of the profile center was non-trivial due to the aforementioned offset between the X-ray peak and the BCG (see Figure \ref{fig:A2495-offset} and Section \ref{sec:offsets}). We decided to evaluate both the BCG and the X-ray peak as centers of the profile and investigate the difference. In Figure \ref{fig:A2495-SBprofBeta} we show the profiles centered on the BCG (left) and the X-ray peak (right), modeled with a single $\beta$-model, while the best fit parameters as well as the chi-square are reported in ~ Table \ref{tab:beta}. 

\begin{figure}[htb]
    \centering
    \includegraphics[scale=0.36]{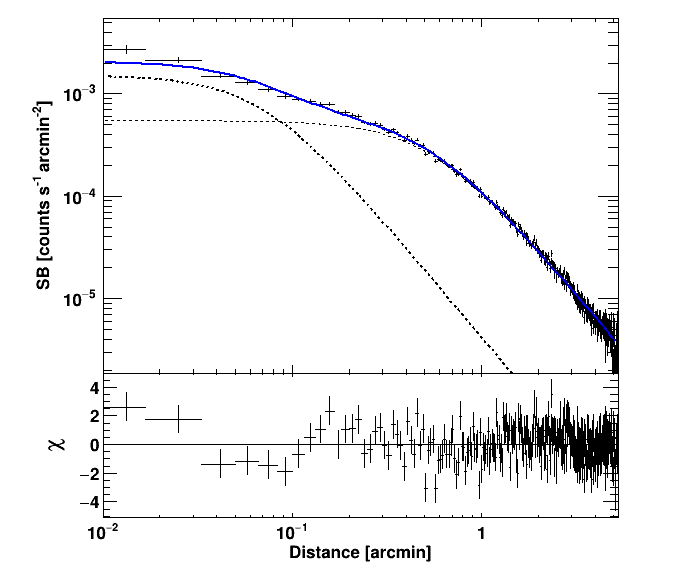}
    \caption{Surface Brightness radial profile centered on the X-ray peak. The blue solid line represents the best-fit double $\beta$-model, and the black dotted lines are the single components of the model.}
    \label{fig:A2495-dubbeta}
\end{figure}

\begin{table}
\centering
\begin{tabular}{|c|c|c|c|c|}
\hline
  Center & $norm$ & $\beta$ & $r_c$[kpc] & $\chi^2$ \\
 \hline
  X & $8.1_{-0.1}^{+0.1}$ & 0.511$_{-0.002}^{+0.002}$ & 35.8$_{-0.6}^{+0.6}$ & 2.53 \\ 
  \hline
  cD & $8.7_{-0.2}^{+0.2}$ & 0.504$_{-0.002}^{+0.002}$ & 33.2$_{-0.6}^{+0.6}$ & 2.59\\
 \hline
\end{tabular}
\caption{Single $\beta$-model best fit parameter (see Figure \ref{fig:A2495-SBprofBeta}, for the surface brightness radial profile). (1) Profile center; (3) model normalization in $10^{-5}$ counts s$^{-1}$ arcmin$^{-2}$; (3) $\beta$-model index; (4) core radius; (5) Reduced chi-square value ($\chi^2$/d.o.f.).}
\label{tab:beta}
\end{table}
It is possible to see from the panels in Figure \ref{fig:A2495-SBprofBeta} that in both cases a significant central emission excess is present (within $\sim15$ kpc from the center). These profiles look very similar, except in the central over-brightness, but these difference can be explained by the different position of the profile center with respect to the X-ray peak: in one case offset (BCG) and the other coincident (X-ray peak). In conclusion, we find that either centering the profile on the BCG or on the X-ray peak does not influences the overall shape of the surface brightness profile and we decided to continue the analysis by centering all profiles and referring all radial quantities to the X-ray peak \citep[see][for comparison]{2022A&A...665A.117C}. 
The presence of a central excess above the single $\beta$-model is consistent with typical surface brightness profiles of cool-core (CC) clusters, that are usually best fitted by the sum of two $\beta$-models. For this reason, we used a double $\beta$-model (with linked $\beta$ parameters, see \citealt{mohr_99}) to take into account the central excess:
\begin{align}
    S(r)=&norm\bigg\{\bigg[1+\bigg(\frac{r}{r_{c1}}\bigg)^2\bigg]^{-3\beta + 0.5} +\\ &+ratio\bigg[1+\bigg(\frac{r}{r_{c2}}\bigg)^2\bigg]^{-3\beta + 0.5}\bigg\} + cost \nonumber
\end{align}
where $norm$ is the normalization of the overall model, $r_{c1,2}$ are the scale radii, $ratio$ is the relative normalization of the inner component, $\beta$ is exponent value of the model and $cost$ is a normalization constant. The best fit parameters for this model are summarized in Table \ref{tab:dbeta}, and the radial profile is shown in Figure \ref{fig:A2495-dubbeta}. This model has a significantly lower $\chi^2$ than the single $\beta$-model, thus providing a better description of the inner profile.

\begin{table*}
\begin{tabular}{|c|c|c|c|c|c|c|}
\hline
Center &$norm$ &$\beta$ & $r_{c1}$[kpc] &  $r_{c2}$[kpc] & ratio & $\chi^2$\\
\hline
X-peak &$5.3_{-0.3}^{+0.3}$ & 0.536$_{-0.003}^{+0.004}$ & 49.1$_{-1.6}^{+1.7}$ & 7.1$_{-0.8}^{+0.9}$ & 2.6$_{-0.2}^{+0.2}$ & 1.21\\
\hline
\end{tabular}
\caption{Double $\beta$-model best fit parameters (see Figure \ref{fig:A2495-dubbeta}, for the radial profile). (1) profile center; (2) model normalization in $10^{-5}$ counts s$^{-1}$ arcmin$^{-2}$; (3) $\beta$-model index; (4) core radius of the inner component; (5) core radius of the external component; (6) normalization factor of the inner component; (7) chi-square value ($\chi^2$/d.o.f.).}
\label{tab:dbeta}
\end{table*}
We also tried to unlink the two beta values but without any statistic improvement, obtaining in both cases a $\chi^2/d.o.f = 1.03$.
\subsection{X-ray Cavities in the ICM}\label{sec:X-cav}
\begin{figure}
    \centering
    \subfigure[]{
    \includegraphics[scale=0.4]{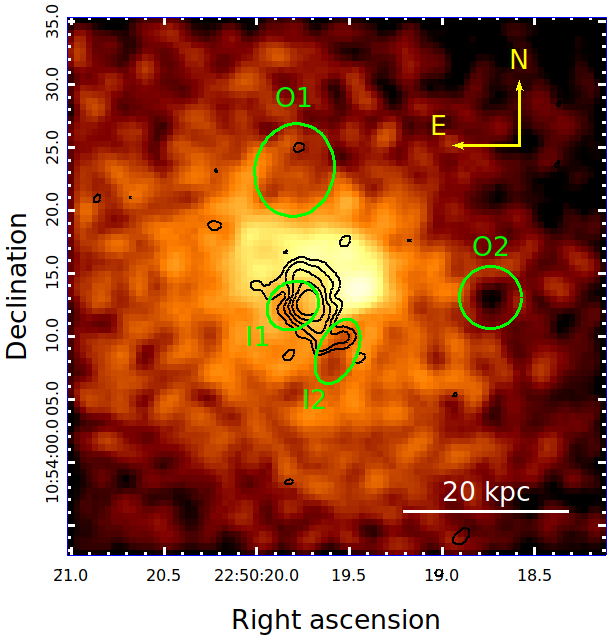}}
    \subfigure[]{
    \includegraphics[scale=0.4]{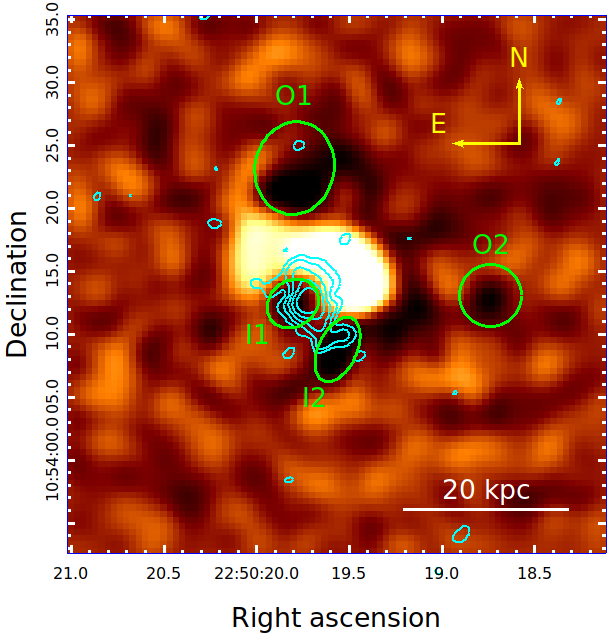}}
    \caption{Panel (a): 0.5-2 keV counts image. Panel (b): 0.5-2 keV unsharp mask made by subtracting a gaussian smoothing of 6" from a 2" one. The cavity regions are highlighted in green and the 5.0 GHz radio contours are reported in black and cyan respectively.}
    \label{fig:A2495-cav}
\end{figure}

One of our main goals is to study the presence of cavities in the central region of this cluster. In the previous work, \citet{pasini_19} found two pairs of depressions in the X-ray emission, corresponding to $\sim30\%$ deficits (at $90\%$ confidence level), with the innermost pair being symmetric with respect to the BCG and the outermost pair being symmetric with respect to the X-ray peak \citep[see][Figure 14]{pasini_19}. However, due to the shallow X-ray observation ($\sim$ 8 ks), the authors were not able to confirm the real nature of these depressions.\\
With our new $Chandra$ observations (120 ks) a much more robust analysis of the significance and size of these cavities is possible. We note that given the reduced {\it Chandra} sensitivity below 1~keV, it is non-trivial to detect low-contrast features even with relatively long exposure. Despite this, the combination of different imaging techniques and detection methods can strengthen the identification of X-ray cavities, as detailed below. To enhance substructures in the cluster core, we created an \emph{unsharp mask image}: we selected the scales that best emphasize structures at the cluster center (2" and 6") and then we subtracted the two version of the convolved 0.5-2 keV with the gaussian kernels at these scales to each other. The resulting image highlights structures between the two scales (Figure \ref{fig:A2495-cav}b). 
\\We inspected the presence of cavities by eyes, both in the 0.5-2 keV counts image and unsharp mask image (see Figure \ref{fig:A2495-cav}), in order to ensure that any identified feature was not a spurious artifact of a single image. On the basis of this visual inspection of the whole cluster core we identified several depressions, and we determined that four of them represent reliable cavities in X-ray surface brightness (see the analysis below). The innermost two depressions (labeled I1 and I2) are located $\sim$5-10 kpc east of the X-ray peak, and appear co-spatial with the radio emission of the BCG. A northern, larger depression (labeled O1) is visible above the bright tail-like structure in the ICM that starting from the X-ray peak extends eastwards. The fourth depression (labeled O2) is located west of the X-ray peak. 

We investigated the significance of each depression with several methods (See appendix for additional details) and in Table \ref{tab:cav} we report their properties. We find that the cavities are detected at just above $2.5\sigma$, and up to $3.8\sigma$ confidence. We also note that the significance of the features depends on the adopted method (see the appendix). This mainly reflects the difficulty of defining a reference surface brightness in a highly-asymmetric ICM. To account for uncertainties in the cavity sizes in the analysis, we assumed a 10$\%$ error on the semi-axes values.
\begin{table}
\centering
\begin{tabular}{|c|c|c|c|c|}
\hline
 Cav & d [kpc] & a [kpc] & b [kpc] & $\sigma$\\
 \hline
    O1 & $15.8\pm1.6$ & $5.6\pm0.6$ & $4.7\pm0.5$ & 3.8\\
 \hline
    O2 & $21.8\pm2.2$ & $3.8\pm0.4$ &$3.8\pm0.4$ & 2.9\\
 \hline
    I1 & $2.0\pm0.2$ & $3.3\pm0.3$ & $2.7\pm0.3$ & 3.3\\
 \hline
    I2 & $6.9\pm0.7$ & $4.1\pm0.4$ & $2.3\pm0.2$ & 2.5\\
 \hline
\end{tabular}
\caption{Size of the detected X-ray cavities, based on the adjacent sector method (see text and Appendix for details). (1) cavity label; (2) distance from the AGN; (3), (4) major and minor semi-axes of each cavity; (5) significance of the depression.}
\label{tab:cav}
\end{table}

\section{Spectral Analysis} \label{sec:an-spec}
We investigated the thermodynamic properties of A2495 through a detailed spectral analysis of the ICM, performed with the software \texttt{XSPEC (v. 12.11.1)}\citep{arnaud_96}. Spectra were fitted in the 0.5-7 keV energy range, using the blanksky files as background. For every thermal model and photoelectric absorption model employed in this work, we used the table of abundances of \citet{asplund_09}.

\subsection{Global properties of the cluster}
To derive the global properties of the cluster, we extracted the spectrum of an ellipse centered on the X-ray peak with semi-major/minor axis of $a=480$ kpc, $b=370$ kpc, and with the same orientation and ellipticity  used for the surface brightness modelling (see Section \ref{sec:SBP}). We fitted the spectrum using a \texttt{tbabs*apec} model, where the galactic absorption was fixed at $N_H=4.41\cdot10^{20} \text{cm}^{-2}$ \citep{HI4PICollaboration_16}; the temperature ($kT$), abundance ($Z$) and normalization ($norm$) of the thermal component were left free to vary, while the redshift was fixed at $z=0.0792$. To measure the unabsorbed flux and luminosity of the thermal emission, we convolved the apec component with the \texttt{cflux} and \texttt{clumin} components. We measured $kT=4.31 \pm 0.05$ keV, $Z=0.60\pm0.03$ $Z_{\odot}$, $F(0.5-7\text{ keV})=1.07\pm0.01\cdot10^{-11}$ erg s$^{-1}$ cm$^{-2}$, $L(0.5-7\text{ keV})=1.66\pm0.01\cdot10^{44}$ erg s$^{-1}$.
These results are consistent to those found by \citet{pasini_19} that are reported here to facilitate the reader: $kT=3.90 \pm 0.20$ keV, $Z=0.60_{-0.10}^{+0.11}$ $Z_{\odot}$, $F(0.5-7\text{ keV})=1.07^{+0.01}_{-0.01}\cdot10^{-11}$ erg s$^{-1}$ cm$^{-2}$, $L(0.5-7\text{ keV})=1.44\pm0.01\cdot10^{44}$ erg s$^{-1}$.
We investigated the presence of an additional thermal component by adding a second \texttt{apec} term. We checked the improvement of the fit quality using the \texttt{ftest}, and this procedure returned a $\text{p-value}=0.83$, showing that added component was not necessary.

\subsection{Projected and Deprojected spectral profiles}\label{sec:pr-dpr_spec}
\begin{figure*}
    \centering
   \subfigure{
    \includegraphics[scale=0.52]{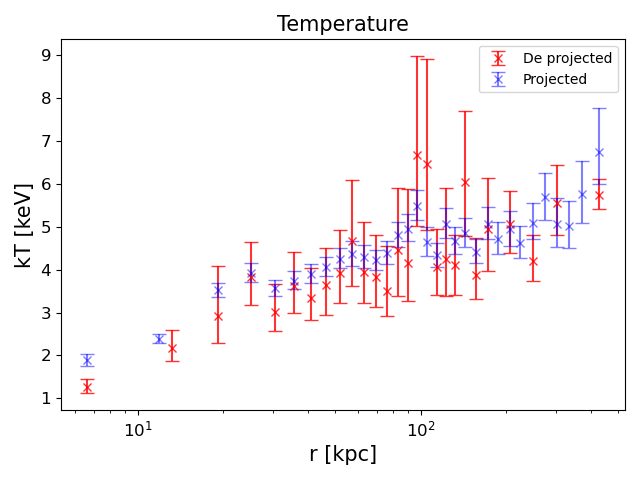}}
    \subfigure{
    \includegraphics[scale=0.52]{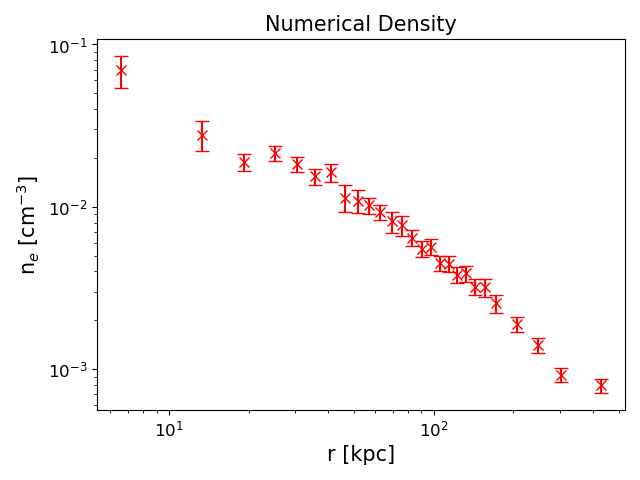}}
    \subfigure{
    \includegraphics[scale=0.52]{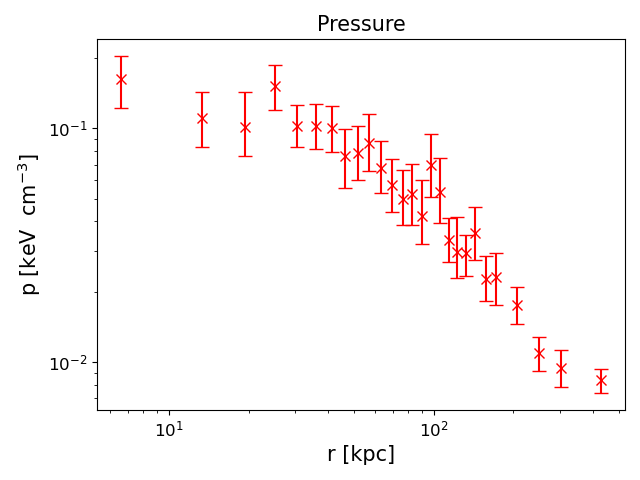}}
    \subfigure{
    \includegraphics[scale=0.52]{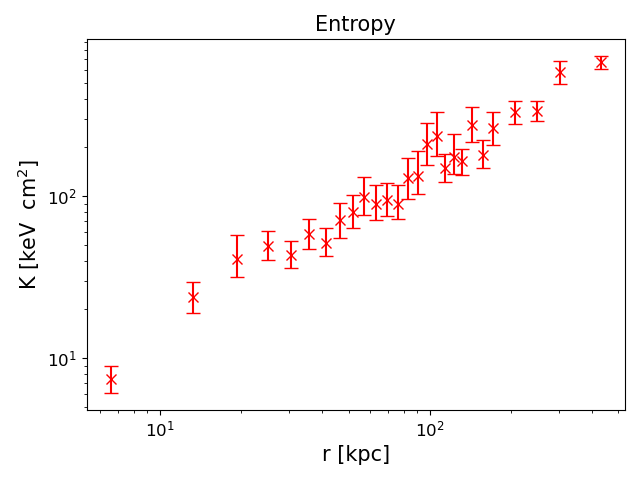}}
    \caption{Deprojected (red) and projected (blue) temperature (\emph{top left}), density (\emph{top right}), pressure (\emph{bottom left}) and entropy (\emph{bottom right}) profiles of the ICM in Abell~2495. The errorbars are at the 1$\sigma$ confidence level.} 
    \label{fig:dprspec}
\end{figure*}

We obtained a projected temperature profile of the ICM by extracting the spectra from concentric elliptical annuli centered on the X-ray peak and extending up to $r\approx 400$ kpc. The bin width was chosen so that each annulus contained a minimum of 3000 net counts. The spectra were fitted with a \texttt{tbabs*apec} model, with fixed hydrogen column density and redshift. The best-fit parameters are reported in Table \ref{tab:projected}, and Figure \ref{fig:dprspec} shows the projected profile of temperature (top left panel, blue profile). Thanks to the high S/N ratio of the observations it was possible to build a high spatial resolution ($\approx5"$) radial profile with relatively small uncertainties (below $10\%$ for the temperature profile).

With the aim of removing the ICM contribution along the line of sight we performed a deprojection of the spectra using the \texttt{projct} component, that is the model became \texttt{projct*tbabs*apec}. This model performs a 3-D to 2-D projection of prolate ellipsoidal shells onto elliptical annuli, by summing the contribute of each shells\footnote{\url{https://heasarc.gsfc.nasa.gov/xanadu/xspec/manual/XSmodelProjct.html}}. For example, the emission from the first annulus has the contribute from all the other shell, the second as well but wihtout the first one and so on and so forth. This procedure is essential to derive the deprojected electron density of the gas. The $norm$ parameter of the \texttt{apec} component is related to the gas density as:
\begin{equation}\label{eq:norm}
    norm=\frac{10^{-14}}{4\pi[D_A(1+z)]^2}\int n_e n_p dV
\end{equation}
where $D_A$ is the angular distance from the source, $z$ is the redshift, $n_e$ and $n_p$ are the electron and proton densities, and $V$ is the projected volume of the emitting region. Assuming $n_e=1.2n_p$ \citep{gitti_12}, it is possible to derive the electron density from Eq. \ref{eq:norm} as:
\begin{equation}
    n_e=\sqrt{10^{14}\bigg(\frac{4\pi\times norm \times [D_A(1+z)^2]}{0.83V}\bigg)}
\end{equation}
where V is the volume of the spherical shells and norm is the normalization of the deprojected \texttt{apec} component. From the electron density and temperature profiles we derived the gas pressure ($p=1.83n_ekT$) and entropy ($K=kT/n_e^{2/3}$). The results of this analysis are reported in Table \ref{tab:deprojected}, and the respective profiles in Figure \ref{fig:dprspec}. These profiles, as the projected temperature one, are consistent with those of a typical cool core cluster. In particular, the entropy and temperature decrease and the density increases towards the center. In the first bin, corresponding to the innermost 10 kpc, the entropy and temperature reach the values of $kT=1.3_{-0.1}^{+0.2}$ keV and $K=7.5_{-1.4}^{+1.5}$ keV cm$^{-3}$.

\subsection{Cooling properties}\label{sec:coolprop}
\begin{figure}
    \centering
    \includegraphics[scale=0.6]{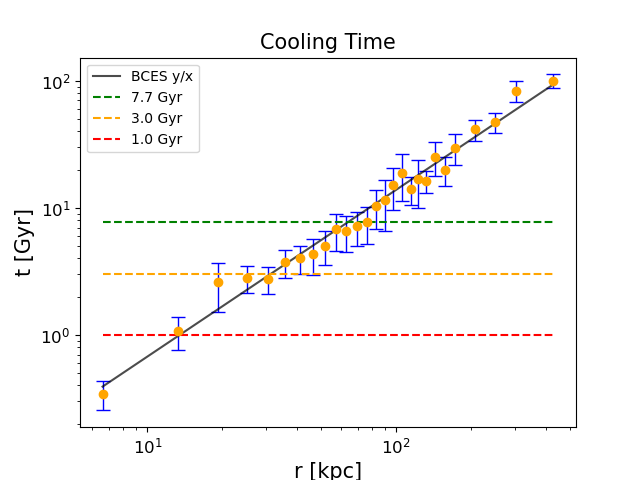}
    \caption{Cooling time profile fitted with a linear regression (black solid line), using the BCES (y/x) method. The age thresholds are also shown with dashed lines: 7.7 Gyr (green, $r_{cool}=63.8\pm0.2$ kpc), 3.0 Gyr (orange, $r_{cool}=29.7\pm0.2$ kpc) and 1.0 Gyr (red, $r_{cool}=14.4\pm0.2$ kpc).}
    \label{fig:tcool}
\end{figure}

\begin{figure}
    \centering
    \includegraphics[scale=0.55]{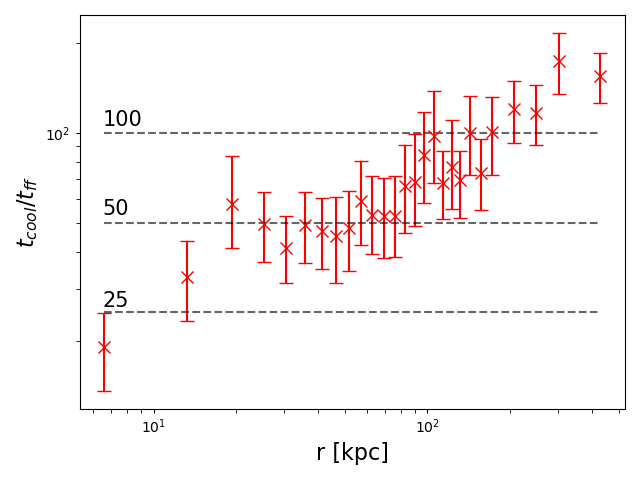}
    \caption{Ratio between the cooling time (eq.\ref{eq:t_cool}) and the free-fall time (eq.\ref{eq:tff}).}
    \label{fig:ti_ratio}
\end{figure}

With the aim of studying the radiative cooling efficiency of the ICM, we used the profiles shown in the previous subsection \ref{sec:pr-dpr_spec} to derive the cooling time radial profile:
\begin{equation}\label{eq:t_cool}
    t_{cool}=\frac{\gamma}{\gamma-1}\frac{kT}{\mu n_e X \Lambda(T)}
\end{equation}
where $\gamma=5/3$, $\mu\approx0.6$, $X=0.7$ and $\Lambda(T)$ are respectively the adiabatic index, the mean molecular weight, the hydrogen fraction and the cooling function \citep{sutherland_93}. Within the so-called cooling radius $r_{cool}$, the $t_{cool}$ falls below the typical relaxation time for a galaxy cluster, considered as the look-back time at $z=1$ ($\approx 7.7$ Gyr). We measured the size of this region by fitting the cooling time profile with a power-law relation, using the \emph{Bivariate Correlated Errors and intrinsic Scatter} \citep[BCES][]{akritas_96} with the Y/X mode and then, selecting the distance at which the fit intersects the age threshold of 7.7 Gyr (see Figure \ref{fig:tcool}). We obtained a cooling radius of $r_{cool}(7.7 \text{ Gyr})=63.8\pm0.2$ kpc. Moreover, we derived the radii within which radiative cooling is even more efficient, i.e. where $t_{cool} <$ 3.0 Gyr and $t_{cool}<1.0$ Gyr. These are respectively $r_{cool}(3.0 \text{ Gyr})=29.7\pm0.2$ kpc and $r_{cool}(1.0 \text{ Gyr})=14.4\pm0.2$ kpc. These results, combined with the central surface brightness excess presented in Section \ref{sec:SBP}, provide a further evidence to classify A2495 as a cool-core cluster.\\
With the cooling time it is also possible to inspect how likely it is for the ICM to form condensed structures and then feed the AGN through chaotic cold accretion \citep[CCA,][]{gaspari_13} . A typical proxy of the ICM ability to condense is the ratio between the cooling time $t_{cool}$ to free fall time $t_{ff}$ defined in section \ref{sec:sloshing}, this is also known as TI ratio. The triggering threshold for the condensation is not well defined, and lies between 8-25 \citep{gaspari_12} up to 70 \citep{valentini_15}. We find that for A2495 this ratio is below 25 within the innermost 15 kpc from the X-ray peak (Figure \ref{fig:ti_ratio}).
It is noteworthy that this region comprises the dust filamentary structure found in the optical HST observation analysed in \citet{pasini_19} (section 4.2).\\
We are unable to use here other criteria like the $t_{cool}/t_{eddy}$ ratio, also known as the \emph{C-ratio} \citep{gaspari_18}, where $t_{eddy}$ is the turbulent turnover time in the hot gas. This is due to the fact that $t_{eddy}$ is often approximated as $L/\sigma_v$, where L and $\sigma_v$ are the injection scale and the 3D velocity dispersion of the hot gas turbulence, respectively. These quantities are not directly measurable for the ICM, but they could be traced by other observables. In particular, the size of the $H\alpha$ emitting nebula or the extent of the region populated by X-ray cavities are proxies for the injection scale, while the width of the $H\alpha$ or other optical emission lines are linked with the $\sigma_v$. This last assumption is appropriate only when the warm gas cooled out recently and still retains information of the turbulent velocity of the hot gas. 
In A2495 the warm gas nebula is compact, regular and shows a smooth velocity gradient \citep{hamer_16}. Such a relaxed nebula is likely to have formed by ICM cooling at least a few $t_{ff}$ ago, and by now it has ''forgotten'' any information about the progenitor hot gas kinematics.\\
\vskip 0.1cm
\noindent To infer more information about the cooling region, we determined the X-ray luminosity within $r_{cool}(t_{cool}<7.7 \text{ Gyr})$. To do so, we extracted the spectrum within $r_{cool}$ and a second spectrum from $r_{cool}$ to the edge of the S3 chip (to perform the deprojection), and we used a \texttt{projct*tbabs*(clumin*apec)} model to fit them.
The lower and upper energy limits of the \texttt{clumin} component were fixed at 0.1 keV and 100 keV, respectively. We obtained $L(<r_{cool})=5.62\pm0.05\times10^{43}$ erg s$^{-1}$. \\Using this value, we estimated the mass deposition rate due to the classical cooling flow process \citep{fabian_94a} as:
\begin{equation}
    \Dot{M}_{cool}=\frac{2}{5}\frac{\mu m_p}{kT}L_{cool}
\end{equation}
where $kT=3.40\pm0.09$~keV is the temperature of the cooling region estimated with a deprojected analysis of the cooling region using a \texttt{projct*tbabs*apec}. We measured a theoretical mass deposition rate of $\Dot{M}_{cool}=47.2\pm0.4$ M$_{\odot}$/yr.\\
We can compare this results with the mass rate inferred directly from the spectra, adding the \texttt{mkcflow} to the thermal model, so that the model becomes \texttt{projct*tbabs*(mkcflow+apec)}. This model describes a cooling flow gas embedded in a hot thermal ambient gas. The \texttt{mkcflow} normalization component represents the mass deposition rate in M$_{\odot}$/yr. We fixed the lower temperature at the lowest value available in \texttt{XSPEC}, that is $kT=0.0808$ keV, while the maximum temperature, as well as the abundance, were tied to those of the \texttt{apec} component. Since no cooling is expected outside $r_{cool}$, we fixed the \texttt{mkcflow} normalization of the outer annulus to zero. With this model we inferred an upper limit on the spectroscopic cooling rate of $\Dot{M}_{cool}<0.3$ M$_{\odot}$/yr, since the fit did not recognize a significant contribution of the \texttt{mkcflow} component to the spectrum. This result was found for other cool core clusters \citep[e.g.][]{peterson_06, pinto_14} and suggests that the effective cooling is at least ten times lower than the one estimated through the classical cooling flow model.
\\Recently, \citet{fabian2022,fabian2023} proposed that significant X-ray cooling rates can be recovered by including in the spectral fits an intrinsic absorption component. \citet{pasini_19} revealed a dust lane in A2495, connected to the BCG and extending along the lower ridge of the H$\alpha$ emission nebula. Thus, we tested whether an absorbed cooling flow model could allow for a higher spectroscopic mass deposition rate (see the details in Appendix \ref{app:hcf}). We extracted the spectrum of the innermost 20~kpc, using a region that encompasses the X-ray peak, the X-ray bright arc-shaped structure, the central AGN, and the dust lane (Fig. \ref{fig:hiddenCF}). Within this region there are 3600 net counts in the 0.5 -- 7 keV band. We fitted the spectrum using the model described in \citet{fabian2022,fabian2023}, finding an absorbed mass deposition rate of $\dot{M}_{abs} = 10.84^{+2.08}_{-2.27}$~M$_{\odot}$/yr, and an intrinsic absorbing column density of $n_{\text{H,int}} = 7.9^{+2.0}_{-1.0}\times10^{21}$~cm$^{-2}$ (with $\chi^{2}/\text{d.o.f.}= 177.06/138$; see Tab. \ref{tab:hiddenCF}). For completeness, we also tested an alternative model. We modified the model used to fit the cooling region (\texttt{tbabs$\ast$(apec + mkcflow)}) by adding an intrinsic absorber to the cooling flow term, thus defining the model: \texttt{tbabs$\ast$(apec + ztbabs$\ast$mkcflow)}. We found an absorbed mass deposition rate of $\dot{M}_{abs} = 11.11^{+2.80}_{-3.15}$~M$_{\odot}$/yr, and an intrinsic absorbing column density of $n_{\text{H,int}} = 3.9^{+0.8}_{-0.7}\times10^{21}$~cm$^{-2}$ (with $\chi^{2}/\text{d.o.f.}= 182.67/139$; see Tab. \ref{tab:hiddenCF}). 
\\The two methods provide fairly consistent results, especially in terms of absorbed mass deposition rate (around 11~M$_{\odot}$/yr). The reduced $\chi^{2}$/d.o.f. of the two models are also comparable (1.29 versus 1.32). The absorbed cooling rate is roughly 25\% of that predicted from the classical cooling flow model ($\sim47$~M$_{\odot}$/yr), and larger than the upper limit of $\dot{M}\leq0.3$~M$_{\odot}$/yr. However, the relatively poor reduced-$\chi^{2}$ of 1.29 and 1.32 limits the reliability of these results. 
\\We conclude that there may be a cooling rate below 1~keV of up to a few M$_{\odot}$/yr in A2495, but the present data do not allow us to draw firm conclusions. Indeed, the \say{hidden cooling flow} model of \citet{fabian2022,fabian2023} has so far been applied to {\it XMM-Newton}/RGS data (with the exception of the very deep Chandra/ACIS data of the outer filament in Perseus, see fig. 5 in \citealt{fabian2022}), that are more suited to detailed spectral tests. Furthermore, any hidden cooling (and absorption) occurring below 1 keV would be hard to constrain with recent {\it Chandra}/ACIS data, that have a reduced sensitivity in this spectral range.

\subsection{Cavities properties}\label{sec:cav-prop}
The lack of a significant mass deposition rate suggests that cooling is counterbalanced by some heating source. We proceeded to evaluate the energy injected by the central AGN during the two outbursts. The energy to excavate cavities can be defined as the thermal energy of the internal gas plus the work done to excavate it through the ICM \citep[e.g.][]{mcnamara_07,mcnamara_12}:
\begin{equation}\label{eq:Ecav}
    E_{cav}=\frac{\gamma}{\gamma-1}pV
\end{equation}
where $\gamma$ is the adiabatic index, $p$ is the pressure of the ICM around the cavity and $V$ is its volume. The plasma within the cavities is assumed to be relativistic, thus $\gamma=4/3$ and the Eq.\ref{eq:Ecav} becomes $E_{cav}=4pV$. The cavity power can be obtained by dividing this energy by the cavity age, that is $P_{cav}=E_{cav}/t_{age}$. To calculate the cavity age, we have considered four methods \citep{birzan_04,ubertosi2021b}:
\begin{itemize}
    \item \emph{Sound Crossing Time.} Assuming the sound speed as the cavity velocity, the age can be defined as $t_{cs}=d/c_s$, where $c_s=\sqrt{\gamma kT/\mu m_p}$ is the sound speed and $d$ is the distance between the cavity center and the AGN.
    \item \emph{Buoyancy Time.} Assuming that the cavity motion is buoyant, its uprise speed is $v_{buo}\approx\sqrt{2Vg/SC}$, where $V$ is the cavity volume, $g$ is the gravity acceleration, $S$ is the cavity area and $C=0.7$ is the drag coefficient. Thus, the cavity age can be computed as $t_{buo}=d/v_{buo}$.
    \item \emph{Refill Time.} It is possible to use the time required to refill the cavity volume by the ICM as a proxy of its age. In this case, $t_{ref}\approx2\sqrt{R/g}$, where $R$ is the cavity radius.
    \item \emph{Expansion Time.} If it is assumed that the cavity is expanding at the sound speed, its age can be defined as $t_{exp}=R/c_s$.
\end{itemize}
It is important to note that $t_{buo}$ and $t_{ref}$ depend on the profile of the gravity acceleration, directly related to the mass profile ($g(r)= \frac{GM(<r)}{r^2}$), which we estimated for Abell~2495 using the hydrostatic mass definition:

\begin{equation}\label{m:idro}
    M_{tot}(<r)=-\frac{k_b T(r) r}{G\mu m_p}\bigg[\frac{\text{dln}\rho}{\text{dln} r} + \frac{\text{dln}T}{\text{dln} r}\bigg] 
\end{equation}

We measured the cavity power of each depression using the above methods (results are reported in Table \ref{tab:pcav}).\\
We also exploited the spectral index map between 1.4 GHz and 5.0 GHz from \citet{pasini_19}, to deduce the radiative age of the electron population inside the cavities. With the available radio data this is possible only for the innermost cavities. Using the spectral index value we measured the radiative age as \citep[see][as other works which used the same method]{eilek_14,bruno_19,ubertosi2021b}:
\begin{equation}\label{eq:spid}
        t_{rad} = \frac{1590\sqrt{B}}{(B^2+B_{CMB}^2)\sqrt{1+z}}\sqrt{\frac{(\alpha-\Gamma)\ln\big({\frac{\nu_2}{\nu_1}}\big)}{\nu_2-\nu_1}}
\end{equation}
where $B_{CMB}=3.25(1+z)^2\text{ [$\mu$G]}$ is the equivalent magnetic field of the Cosmic Microwave Background (CMB), $B=B_{CMB}/\sqrt{3}\text{ [$\mu$G]}$ is the magnetic field that minimizes the energy loss in the cavity regions (maximizing the ages), $\Gamma=0.7$ is the injection index (that ranges between 0.5 -- 0.9, see \citealt{biava_21} and reference therein), and $\nu_1$ and $\nu_2$ are 1.4 GHz and 5 GHz, respectively.
It should be noted that the age calculated using this method are to be considered as an upper limit for the cavity age.
The age of the cavities derived from the X-ray methods and the radio method are summarized in Table \ref{tab:tcav}. We also compute the average between the ages describe above, excluding the radiative one. For these mean times ($t_{mean}$), as well as for other results showed below, we report the dispersion $\sqrt{\frac{\Sigma (<q>-q_i)^2}{N-1}}$, in order to provide an estimate of the systematic uncertainty given by the combination of values calculated with different methods.
\\As expected, the internal cavities result younger than the external ones. We also note that the upper limit from the radio data is consistent with the X-ray results.
\begin{table*}

\renewcommand{\arraystretch}{1.3}
\begin{tabular}{|c|c|c|c|c|c|c|}
\hline
Cav. & $t_{cs}$ & $t_{buo}$ & $t_{ref}$ & $t_{exp}$& $t_{rad}$ & $t_{mean}$ $(1\sigma)$  \\
\hline
\rule[-2mm]{0mm}{3mm}
O1 & 18.1 $\pm$ 3.7 & 46.7 $\pm$ 11.0 & 55.1 $\pm$ 8.9 & 5.9 $\pm$ 0.6 & - & 31.5 $\pm$ 3.7 ($\pm$23.3) \\
\hline
\rule[-2mm]{0mm}{3mm}
O2 & 24.9 $\pm$ 5.2 & 71.7 $\pm$ 17.0 & 47.1 $\pm$ 7.6 & 4.3 $\pm$ 0.5 & - & 37.0 $\pm$ 4.8 ($\pm$29.0)\\
\hline
\rule[-2mm]{0mm}{3mm}
I1 & 2.6 $\pm$ 0.2 & 7.1 $\pm$ 1.7 & 38.1 $\pm$ 6.2 & 3.9 $\pm$ 0.4 & $<$62 & 12.9 $\pm$ 1.6 ($\pm$16.9)\\
\hline
\rule[-2mm]{0mm}{3mm}
I2 & 9.1 $\pm$ 0.8 & 26.5 $\pm$ 6.3 & 39.4 $\pm$ 6.3 & 4.2 $\pm$ 0.4 &$<$89 & 19.8 $\pm$ 2.2 ($\pm$16.2)\\  
\hline
\rule[-2mm]{0mm}{3mm}
OUTER & 21.5 $\pm$ 3.2 & 59.2 $\pm$ 10.1 & 51.1 $\pm$ 5.9 & 5.1 $\pm$ 0.4 & - & 34.3 $\pm$ 3.0 ($\pm$18.6)\\
\hline
\rule[-2mm]{0mm}{3mm}
INNER & 5.9 $\pm$ 0.4 & 16.8 $\pm$ 3.3 & 38.8 $\pm$ 4.4 & 4.1 $\pm$ 0.3 & $<$ 75.5 & 16.4 $\pm$ 1.4 ($\pm$11.7)\\
\hline
\end{tabular}
\caption{Cavity ages in Myr. $t_{cs}$ sound crossing time; $t_{buo}$ buoyancy time; $t_{ref}$ refill time; $t_{exp}$ expansion time; $t_{rad}$ radiative age; $t_{mean}$ mean age value of the X-ray methods. OUTER and INNER are respectively the mean age for each pair of cavities and for each type of age (without $t_{rad}$). The value of the 1$\sigma$ dispersion are reported between parenthesis for $t_{mean}$. }
\label{tab:tcav}
\end{table*}

\begin{table*}

\renewcommand{\arraystretch}{1.5}

\begin{tabular}{|c|c|c|c|c|c|}
\hline
Cav. & $P_{cs}$ & $P_{buo}$  & $P_{ref}$  & $P_{exp}$   &$P_{mean}$ $(1\sigma)$\\
\hline
\rule[-2mm]{0mm}{3mm}
O1 & $18.1\pm 10.0 $ & $7.0\pm 3.9$ & $6.0\pm 3.1$ & $56.0\pm 28.0$ & $21.8 \pm 7.5$ ($\pm$23.5) \\
\hline
\rule[-2mm]{0mm}{3mm}
O2 & $5.9 \pm 3.2$ & $2.0\pm 1.1$ & $3.1\pm 1.6$ & $34.0 \pm 18.0$ & $11.3 \pm 4.6$ ($\pm$15.3)\\
\hline
\rule[-2mm]{0mm}{3mm}
I1 & $27.0\pm 11.1$ & $9.9\pm 4.7$ & $1.8\pm 0.8$& $18.0 \pm 8.0 $ & $14.2 \pm 3.6$ ($\pm$10.8)\\
\hline
\rule[-2mm]{0mm}{3mm}
I2 & $7.0\pm 3.0$ & $2.4\pm 1.1$ & $1.6\pm 0.7$ & $15.0 \pm 6.0$ & $6.5 \pm 1.7$ ($\pm$6.1)\\
\hline
\rule[-2mm]{0mm}{3mm}
OUTER & $24.0 \pm 10.5$ & $9.0\pm 4.1$ & $9.1\pm 3.5$ & $90.0 \pm 33.3$ & $33.1 \pm 8.3$ ($\pm$28.0)\\
\hline
\rule[-2mm]{0mm}{3mm}
INNER & $34.0 \pm 11.5$ & $12.3\pm 4.8$ & $3.4\pm 1.1$ & $33.0\pm10.0$ &$20.7 \pm 4.0$ ($\pm$12.4)\\
\hline
\rule[-2mm]{0mm}{3mm}
TOT & $58.0 \pm 15.5$ & $21.3\pm 6.3$ & $12.5\pm 3.6$ & $123.0 \pm 34.8$ & $53.8 \pm 9.7$ ($\pm$30.7)\\
\hline
\end{tabular}
\caption{Powers of the X-ray cavities in units of $10^{42}$ erg s$^{-1}$. $P_{cs}$, power obtained using the sound crossing time $t_{cs}$; $P_{buo}$, power obtained using the buoyancy time $t_{buo}$; $P_{ref}$, power obtained using the refill time $t_{ref}$; $P_{exp}$, power obtained using the expansion time $t_{exp}$; $P_{mean}$, mean power value for each cavity and the 1$\sigma$ dispersion are reported between parenthesis. OUTER and INNER, sum of the cavity powers for each pair of cavities. TOT, sum of the all cavity powers.}
\label{tab:pcav}
\end{table*}

\subsection{Analysis of the possible cold front}\label{sez:cf}
\begin{figure*}
    \centering
    \subfigure[]{
    \includegraphics[scale=0.4]{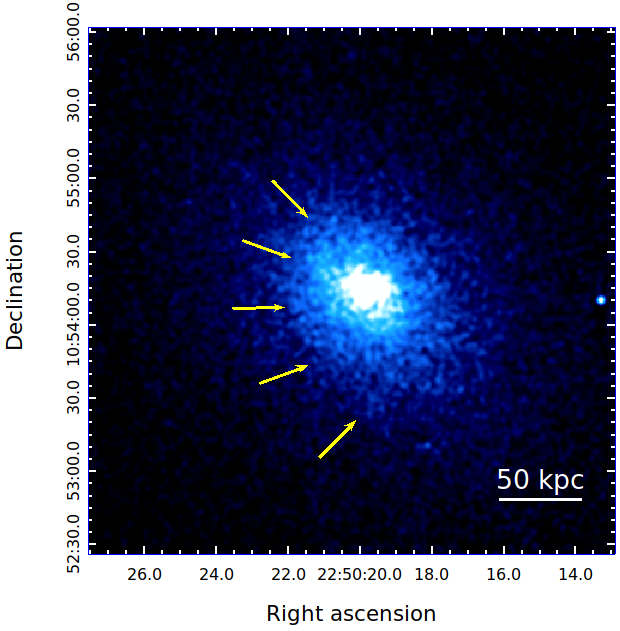}}
    \subfigure[]{
    \includegraphics[scale=0.4]{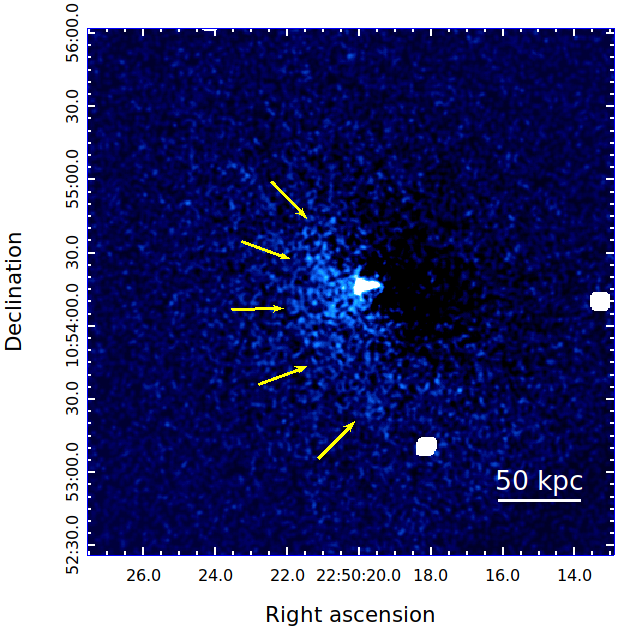}}
     \subfigure[]{
    \includegraphics[scale=0.28]{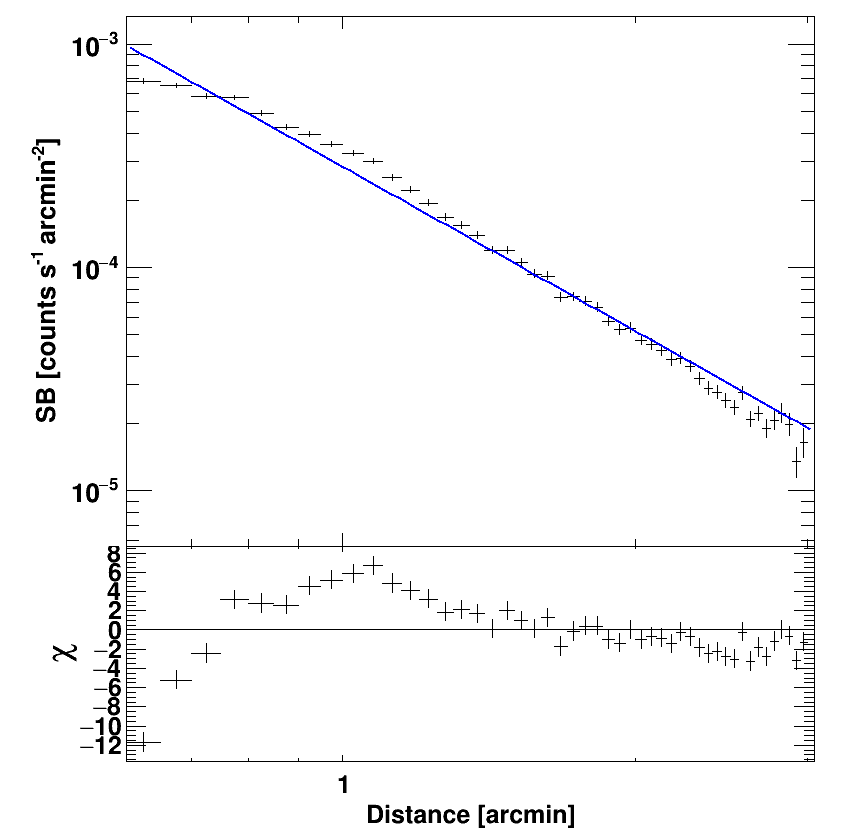}}
    \subfigure[]{
    \includegraphics[scale=0.28]{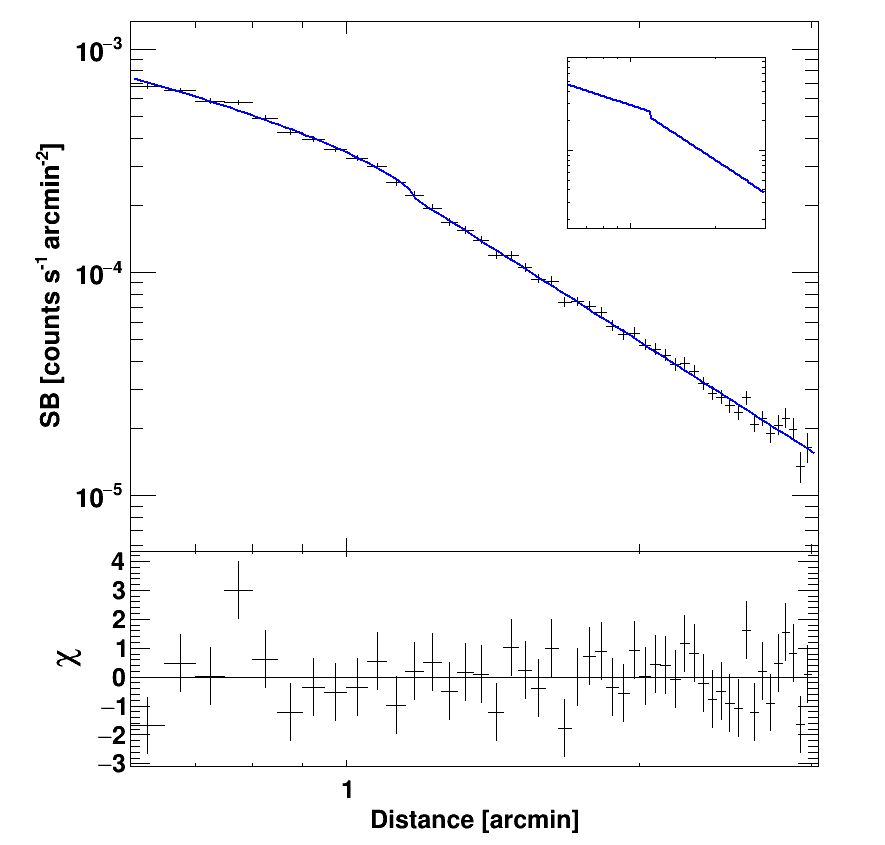}}
    \caption{(a)-(b) 0.5-2 keV image and residual image (obtained by subtracting the best fit double $\beta$-model from the image) of the cluster center, the yellow arrows indicate the shape of the putative discontinuity. The white regions in the residual image are point sources masked to avoid any contamination in the analysis. (c) Surface brightness profile across the discontinuity modeled using a single power law. (d) Same profile modeled using a broken power law. The residuals of each fit are shown in the bottom panel respectively.}
    \label{fig:CF}
\end{figure*}

\begin{figure}
    \centering
    \subfigure[]{
    \includegraphics[scale=0.4]{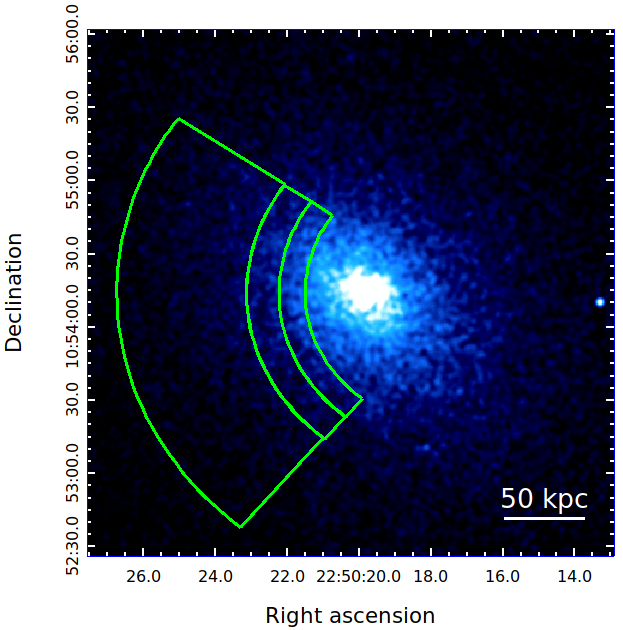}}
    \subfigure[]{
    \includegraphics[scale=0.5]{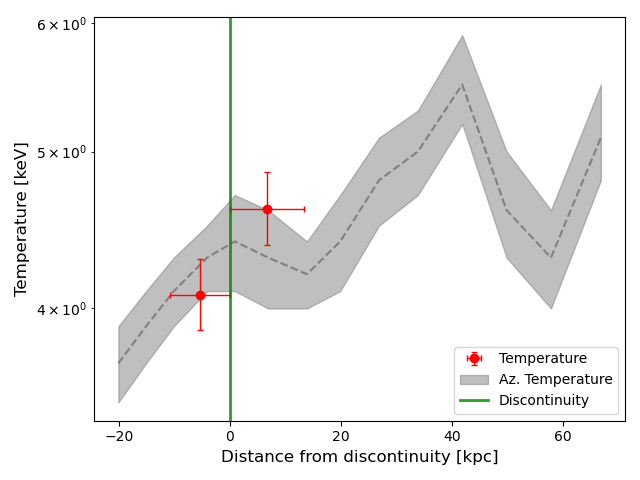}}
    \subfigure[]{
    \includegraphics[scale=0.5]{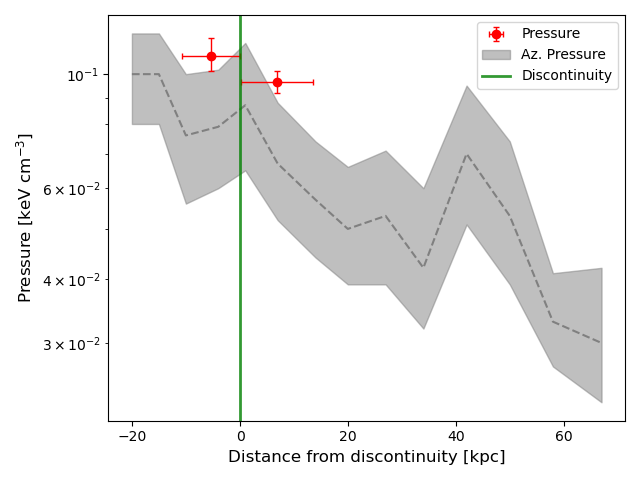}}
    \caption{(a) 0.5-2 keV image with the region used for the spectral analysis around the discontinuity. (b-c) Projected temperature and de-projected pressure values across the discontinuity with the azimuthally average values in shaded gray as reference.}
    \label{fig:cf-imgspec}
\end{figure}

To further characterize the dynamical state of the cluster, we investigated the presence of surface brightness discontinuities in the ICM. We performed this search in \texttt{PROFFIT}, by extracting surface brightness profiles from elliptical sectors which encompass the areas where there is most likely to be a discontinuity, by visual inspection of the residual image (Figure \ref{fig:CF}b). The resulting background-subtracted, exposure corrected surface brightness profiles were fitted using both a single and broken power law.
The analytical form  for the single power law is:
\begin{equation}
    S(b)=norm\bigg(\frac{b}{b_s}\bigg)^{-\beta}
\end{equation}
where $\beta$ is the slope, $b_s$ is a scale radius, and $norm$ is the normalization. To describe a discontinuity in the surface brightness we use a broken power law with a density jump, projected along the line of sight $l$ \citep{eckert_16, Owers_09}:
\begin{equation}
    S(b)=norm \int F(r)^2dl   \text{,  with $r^2=b^2+l^2$}
\end{equation}
where the integration is performed along the line of site.  $F(r)$ is the 3D density distribution, defined as:
\begin{equation}\label{eq:bknpow}
    F(r)= \bigg\{\begin{array}{lr}
       r^{-\alpha_1}  & \text{if } r < \text{cutrad}  \\
       \frac{1}{\text{jump}}r^{-\alpha_2}  & \text{otherwise}
    \end{array}
\end{equation}
where $\alpha_{1,2}$ are the two slopes, $cutrad$ is the jump position along the radial profile (i.e. the distance at which the slope changes), $jump$ is the density jump amplitude. We considered a discontinuity as "detected" if the broken power law represented a better fit to the profile with respect to the single power law. Only one discontinuity was found in A2495, at 58$^{+0.9}_{-4.2}$ kpc east of the center, approximately between 145° and 220° counter-clockwise with respect to the declination axis. This feature is appreciable both in the 0.5-2 keV image and in the residual image (Figure \ref{fig:CF}a,b).
Across this region, the surface brightness profile shows a shallow but appreciable jump ($jump=1.14\pm0.02$ from the best fit model in eq. \ref{eq:bknpow}), where the broken power law is the best fit model instead of a single one with a confidence level above $5\sigma$ ($\chi^2/d.o.f=0.97$ for the broken powerlaw and $\chi^2/d.o.f=10.34$ for the single one). The parameters for both single and broken power law are reported in Tables \ref{tab:bpcf}, while their fit in Figure \ref{fig:CF}c,d.
\begin{table*}
        %\centering
        \renewcommand{\arraystretch}{1.3}
        \begin{tabular}{|c|c|c|c|c|c|c|}
        \multicolumn{6}{c}{\textbf{Power law}}\\
        \hline
        \multicolumn{2}{|c|}{
        $\beta$} & $r_{s}$ [arcmin] & \multicolumn{2}{c|}{$norm$ [cts s$^{-1}$ arcmin$^{-2}$]} & $\chi^2$ & $d.o.f$ \\
        \hline
        \multicolumn{2}{|c|}{$2.45 \pm 0.01$} & $0.69$ & \multicolumn{2}{c|}{$1.7\cdot10^{-4}$} & 465.3 & 45\\
        \hline
        \multicolumn{6}{c}{\textbf{Broken power law}}\\
        \hline
        $\alpha_1$ & $\alpha_2$ & $cutrad$ [arcmin] & $norm$ [cts s$^{-1}$ arcmin$^{-2}$] & $jump$ & $\chi^2$ & $d.o.f$ \\
        \hline
        $0.96 \pm 0.05$ & $1.89 \pm 0.02$ &  0.64$^{+0.01}_{-0.05}$ & $2.9_{-0.7}^{+0.1}\cdot10^{-4}$ & $1.14 \pm 0.02$ & 41.9 & 43\\
        \hline
        \end{tabular}
        \caption{\emph{Top panel}: single power law best fit parameter. $\beta$ Power law slope, $r_s$ scale radius, $norm$ power law normalization. The uncertainties on $r_s$ and $norm$ are not reported because the fit was not able to constraint them at 1$\sigma$ confidence. \emph{Bottom panel}: broken power law best fit parameter $\alpha_{1,2}$ power law slopes, $cutrad$ jump position relative to the X-ray peak, $norm$ model normalization, $jump$ jump magnitude. For the both panels are shown also the chi-squares ($\chi^2$) and the degrees of freedom ($d.o.f$). }
        \label{tab:bpcf}
\end{table*}
We also performed a blind inspection with sectors of 90° and 45° all around the cluster center, but without any additional findings.
\\We measured the temperature across the discontinuity in order to determine its nature. For this spectral fit we chose the bin width of the sectors inside and outside the interface so that they had at least 3000 counts each, and we used a \texttt{tbabs*apec} model. We obtained a temperature value of $4.08_{-0.20}^{+0.21}$ keV inside, and $4.61_{-0.23}^{+0.25}$ keV outside. In Figure \ref{fig:cf-imgspec}b these values are compared with the azimuthally averaged profile. Moreover, we investigated the pressure values through the edge by performing a deprojected analysis, adding a third external region to use the \texttt{projct*tbabs*apec} model (Figure \ref{fig:cf-imgspec}a). With this analysis we obtained a temperature value of $3.55_{-0.44}^{+0.54}$ keV and $4.44_{-0.39}^{+0.44}$ keV inside and outside the edge, respectively. For the pressure, the values found are $1.08_{-0.07}^{+0.09}\cdot 10^{-1}$ keV cm$^{-3}$ and $0.96_{-0.05}^{+0.05}\cdot 10^{-1}$ keV cm$^{-3}$ inside and outside the edge.
This spectral study is not conclusive, but suggests that the ICM is cooler inside the discontinuity than outside, while the pressure is nearly constant through the interface (Figure \ref{fig:cf-imgspec}c). Thus, we classify the detected edge as a probable cold front.
Together with the evidence that the broken power law provides the best fit to the surface brightness profile, and the morphology of the spectral maps shown in the next paragraph, it is likely that the sloshing mechanism is acting in this galaxy cluster.

\subsubsection{Spectral maps of the ICM}\label{sez:specmap}
\begin{figure*}
    \centering
    \includegraphics[width=\linewidth]{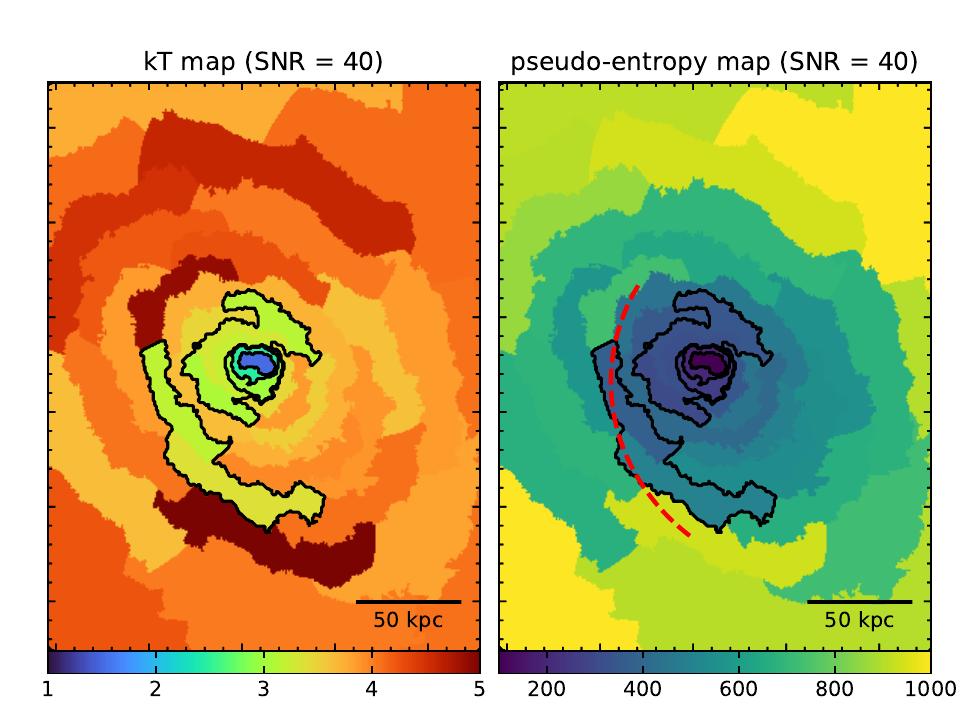}
    \caption{Spectral maps of temperature $kT$ (left panel, in keV) and pseudo-entropy (right panel, arbitrary units) obtained by setting the minimum SNR during binning to 40. The relative uncertainties vary between 10\% -- 12\% for the temperature map and between 14\% -- 16\% for the pseudo-entropy map. In both panels, black contours trace the regions where the ICM temperature is lower than 3.4 keV. In the {\it right panel}, the red dashed arc corresponds to the cold front in Abell~2495 (see Sec. \ref{sez:cf} for details).}
    \label{fig:SNR40}
\end{figure*}

To inspect the 2D distribution of ICM temperature and entropy, we generated spectral maps by binning the 0.5 - 7 keV image of Abell~2495 with the \texttt{CONTOUR BINNING} algorithm \citep{sanders_06}, and then fitting the spectrum of each region with a \texttt{tbabs$\ast$apec} model, fixing the redshift of the cluster and leaving the abundance free to vary. The bin size is set by the minimum signal-to-noise ratio (SNR) that we wish to achieve. For the study of the temperature and entropy distribution, we selected a minimum $\text{SNR} = 40$, that allows us to obtain relative uncertainties of 10\% -- 12\% on kT while preserving a good spatial resolution. The (pseudo-)entropy was derived from the best-fit temperature and normalization ($norm$) as $kT\,\times (norm/n_{pix})^{-1/3}$, where $n_{pix}$ is the number of pixels in each bin (see e.g., \citealt{ubertosi_23}).
\\ The contours overlaid on Figure \ref{fig:SNR40} trace the bins of the temperature map with a temperature $kT\leq3.4$~keV, that is the temperature of the cooling region (\ref{sec:coolprop}). The contours emphasize that the cooler ICM is preferentially found at the center (as expected), and along a tail-like feature on the south-east side of the cluster (corresponding to the positive excess in the residual image of Figure \ref{fig:CF}b). This configuration is reminiscent of the ICM spiral morphology commonly found in sloshing clusters (see e.g., \citealt{ghizzardi_14}). Furthermore, we show in Figure \ref{fig:SNR40} (right panel) that the cold front identified in Section \ref{sez:cf} approximately traces this tail-like cool feature, further supporting its sloshing origin.

\section{Discussion} \label{sec:disc}
\subsection{Heating -- cooling balance}\label{sez:heat-cool}
\begin{figure}
    \centering
    \includegraphics[scale=0.6]{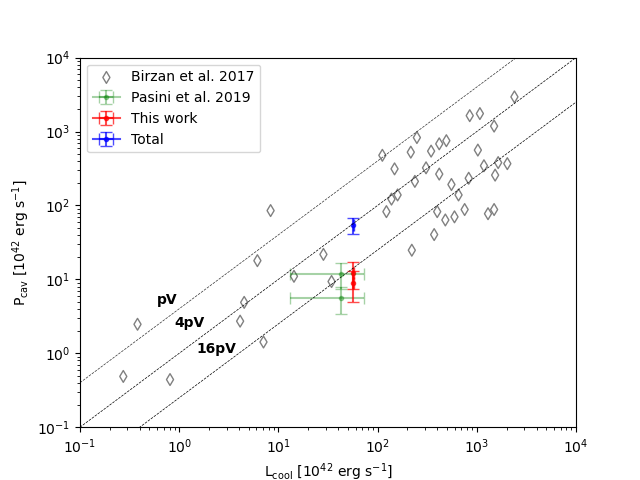}
    \caption{Cavity power versus X-ray luminosity within $r_{cool}$: the gray rhombuses are the result extracted from \citet{birzan_17}, the green and red dots are the results from the \citet{pasini_19} and this work, respectively. The blue dot represents the sum of the power of the two cavity pairs.}
    \label{fig:pcavlcool}
\end{figure}

To understand the heating -- cooling balance in A2495, we compare the cavity power with the X-ray luminosity estimated within the cooling radius. We consider the power calculated with the expansion time for the internal pair of cavities and the power calculated with the average of the dynamic times ($t_{cs}$, $t_{buo}$, $t_{ref}$) for the external one. We used this assumption because the inner cavities I1 and I2 are thought to be still expanding, while the outer ones, O1 and O2, are more likely to be in their buoyancy phase. With these considerations, the power for each pair of cavities are the following:
\begin{equation*}
    \begin{array}{cc}
         \text{OUTER} & P_{cav}=1.4\pm 0.3 (\pm0.9) \cdot 10^{43} \text{erg s$^{-1}$}  \\
         \text{INNER} & P_{cav}=3.3\pm1.0 \cdot 10^{43} \text{erg s$^{-1}$}
    \end{array}
\end{equation*}
As done before, the 1$\sigma$ dispersion is reported between parethensis for the outer P$_{cav}$. Since the cavity ages are always smaller than the cooling time at their distance from the center ($t_{cav}\approx10^7$ yr and $t_{cool}\approx10^9$ yr), we estimated the total heating power by summing the powers of the two cavity pairs. Thus, the total power provided by the cavities is $P_{cav}=4.7\pm1.3 \cdot 10^{43}$ erg s$^{-1}$. This value is consistent with the radiative losses within the cooling region, given by $L_{cool}=5.7\pm0.1 \cdot 10^{43}$ erg s$^{-1}$, suggesting that AGN feedback can compensate for the radiative cooling in this cluster.  
\\To place our results in the context of other cool core clusters, we add A2495 to the plot of the $P_{cav}$-$L_{cool}$ relation (Figure \ref{fig:pcavlcool}). We select the cavity power obtained using the $t_{buo}$ ages, since this is the typical timescales used for cavity sample studies \citep[e.g.][]{birzan_17}. This plot shows that the results of our work are consistent with the previous ones of \citet{pasini_19}, but with a significant improvement of the accuracy in the $L_{cool}$, thanks to the high statistic obtained with the new $Chandra$ observations. Additionally, we confirm that A2495 follows the general trend for cool core clusters found in \citet{birzan_17,birzan_12}. 

\subsection{Interplay between AGN Feedback and offsets}
We investigated two possible scenarios that may explain the presence and the interplay between the offsets and the imprints of AGN feedback (i.e. the cavities):
\begin{itemize}
    \item[i)] {\emph{Mass Uplift}}: In this case, the formation and the buoyant rise of the cavities are the main driver of the offsets.
    \item[ii)]{\emph{Sloshing}}: In this case, both the two cavity generations and the offsets are the result of the ICM oscillation (or sloshing) around the center of the potential well due to a previous dynamic interaction of the cluster with another one (e.g. minor merger). 
\end{itemize}

\subsubsection{Mass uplift}\label{sec:uplift}
\begin{figure}
    \centering
    \includegraphics[scale=0.32]{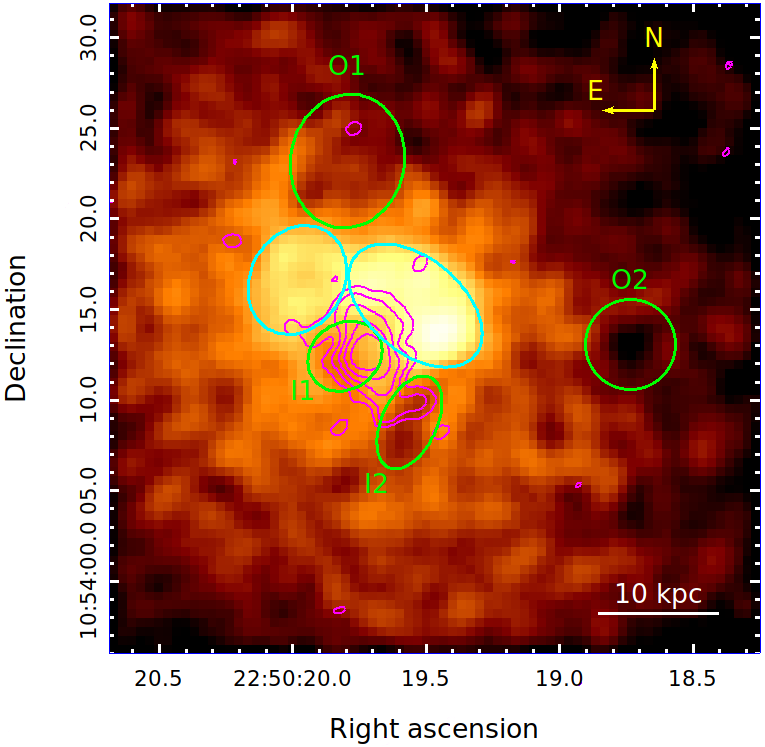}
    \caption{0.5-2 keV $Chandra$ image of the cluster center. The cavity regions are reported in green, the regions enclosing the X-ray peak in cyan. The contours in magenta are the 5.0 GHz radio contours.}
    \label{fig:uplift}
\end{figure}

As already explain in Section \ref{sec:intro}, the cavities formation and rise can push out a fraction of the cooling gas from the bottom of the gravitational potential well. In order to understand if this is the case for A2495, we verified if the cavities are able to move an amount of mass comparable with that of the X-ray peak. To obtain this information, we compared the mass of the X-ray peak with that which would be contained in the regions of the cavity (see Figure \ref{fig:uplift}). The masses were calculated as $m=\rho V$, where $\rho$ is the density at the distance of the region, obtained through the spectral analysis, and $V=\frac{4}{3}\pi ab^2$ is the volume. To compute the mass of the X-ray peak we considered two ellipses, the sum of which encloses most of the peak bright emission (cyan regions in Figure \ref{fig:uplift}). The density of the innermost ellipse is assumed to be equal to the first value of the density radial profile (see Table \ref{tab:deprojected}), this density is also adopted for the more external one (to the east) but is re-scaled by the square root ratio of their brightness ($n\propto\sqrt{I}$).
\\ We obtained the following masses:
\begin{itemize}
    \item[-] X-ray peak: $M=1.3 \pm 0.1 \cdot 10^9$ M$_{\odot}$.
    \item[-] Internal cavities: $M=1.4\pm0.3\cdot10^8$ M$_{\odot}$.
    \item[-] External cavities: $M=3.8\pm0.7\cdot10^8$ M$_{\odot}$.
    \item[-] Internal + External: $M=5.2\pm0.8\cdot10^8$ M$_{\odot}$.
\end{itemize}
These values suggest that the cavities cannot move an amount of mass consistent with that contained in the X-ray peak.
This evidence points to this scenario being implausible or at least not likely. Moreover, the cavity positions seem not consistent with uplift, and it would be still very difficult to hydrodynamically uplift such a diffuse medium with a strongly bipolar and localized injection of bubbles.

\subsubsection{Sloshing of the Intracluster Medium}\label{sec:sloshing}
The offsets between the ICM phases and the BCG suggest that the gas in the innermost region is not in hydrostatic equilibrium. Coupled with the investigation of cold fronts in the ICM (see Section \ref{sez:cf}), it is likely that sloshing is shaping the X-ray morphology of Abell~2495, and may in turn be responsible for such offsets \citep[see][for a comprehensive reviews]{zuhone_16,zuhone_16b,zuhone_19}. Sloshing might control the frequency of the AGN feedback process by inducing a relative motion between the BCG and the cold or cooling ISM/ICM. This could intermittently reduce the fuel available to the black hole to start the feedback cycle. When the dense gas returns towards the center, the AGN can restart its activity, creating a new generation of cavities. It is therefore interesting to constrain the timescales which regulate the sloshing-induced dynamics.
\\\citet{pasini_19} already proposed this scenario for A2495, and they verified its likelihood by comparing the characteristic oscillation period of the ICM at the cavities scale with the difference in the ages of the two pair of putative cavities.
We performed a similar analysis with more accurate estimates of the radial profiles of the ICM properties, as well as geometry and ages of the confirmed cavities. Two dynamical times are relevant for the gas dynamics:

\begin{itemize}
    \item \emph{Free fall time} This is a lower limit for the relative oscillation period of the dense gas-BCG system:
    \begin{equation}\label{eq:tff}
        t_{ff}=\sqrt{\frac{2r^3}{GM(r)}}
    \end{equation}
    
    \item \emph{Brunt-V\"ais\"al\"a time} This is the buoyancy oscillation period in a stable, stratified atmosphere. It derives by the Brunt-V\"ais\"al\"a frequency:
    \begin{equation}\label{eq:fbv}
        \omega_{BV}(r)=\sqrt{\frac{3GM(r)}{5r^3}\frac{d\text{ln} K}{d\text{ln} r}}
    \end{equation}
    which leads to a timescale $t_{BV}=\frac{2\pi}{\omega_{BV}}$
\end{itemize}

The mass profile was calculated using the hydrostatic mass (see Section \ref{sec:cav-prop}) and the logarithmic derivative of the entropy was derived using the BCES library, by fitting the logarithmic entropy profile (Figure \ref{fig:dprspec} bottom right panel). From the fitting process we obtained the following best fit regression:
\begin{equation*}
    log(K) = (1.01\pm0.05)log(r)+(-69.82\pm2.46) \text{ [keV cm$^2$]}
\end{equation*}

\begin{figure}
    \centering
    \includegraphics[scale=0.5]{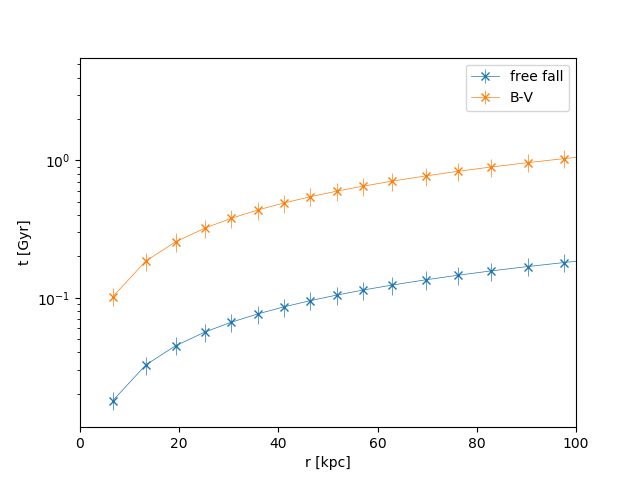}
    \caption{Free fall (blue) and Brunt-V\"ais\"al\"a (orange) times for the inner 100 kpc. It is important to note that the cavity's scale is within the first point ($\lesssim 10$ kpc)}
    \label{fig:tffbv}
\end{figure}

We calculated both timescales at each radius and the resulting profiles are shown in Figure \ref{fig:tffbv}.

As one can see from the plot, the free fall time is always smaller than the BV time at least of a factor of 6 \citep[see][]{su_17,ubertosi2021a}. In the innermost radius, where the cavities reside, the two times are:

\begin{equation*}
\begin{array}{l}
    t_{ff}=17.9 \pm 2.7 \text{ Myr}  \\
    t_{BV}=104.2 \pm 15.6 \text{ Myr} 
\end{array}
\end{equation*}

In the absence of detailed kinematic information, we assumed as a rough sloshing timescale the average of these two times, with the dispersion as its uncertainty, thus $t_{sloshing}=61.1 \pm 15.8 (\pm 43.2) \text{ Myr}$. We compare this time with the difference in the ages of the two pairs of cavities. We adopt the mean of the dynamic ages ($t_{cs}$, $t_{bou}$, $t_{ref}$) for the external pair and the expansion time for the internal one, following the same reasons exposed in section \ref{sec:disc}:

\begin{equation*}
    \begin{array}{l}
         t_{sloshing}=61.1\pm 15.8\,(\pm43.2) \text{ Myr} \\
         \Delta t_{cav}=<t_{dyn}^{OUT}>-t_{exp}^{IN}=39.8\pm10.7\,(\pm19.8) \text{ Myr}
    \end{array}
\end{equation*}

Again, the dispersion between the two dynamical ages is a measure of the uncertainty for the age difference.\\
This result shows that the assumed sloshing time is consistent within uncertainties with the difference in the ages of the two pairs of cavities. A possible scenario we propose is that sloshing could regulate the timescales of the AGN feedback. If, instead, the sloshing motion is characterized by $t_{BV}$ \citep[e.g.][]{su_17}, the timescales relation $\Delta t_{cav} < t_{sloshing}$ would suggest that ICM cooling occurs in (at least) the whole region encompassing the three offsets, not just in the current X-ray peak. This is certainly possible, as indicated by the timescale ratio $t_{cool}/t_{ff} \lesssim 25$ for $r \lesssim 15$ kpc (Section \ref{sec:coolprop}).\\
Future theoretical and observational insights on A2495 and on the sloshing mechanism may provide additional clues to identify the correct scenario.

\section{Summary and Conclusions} \label{sec:concl}
In this paper we presented an investigation of the galaxy cluster A2495 using new, deep ($\sim$ 130 ks) $Chandra$ observations. Through an accurate morphological and spectral analysis of the data we obtained the following results:

\begin{itemize}
    \item We confirmed the presence of a triple spatial offset in this cluster between the BCG, the X-ray peak of the ICM, and the warm gas peak. Specifically, the X-ray peak is located $\sim$6.2 kpc away from the BCG and $\sim$3.9 kpc away from the H$\alpha$ emitting gas. We secured the cool-core nature of this cluster both from the surface brightness profile (Section \ref{sec:an-morph}), that is well described by a double-$\beta$ model, and from the deprojected radial spectral analysis, which reveals that the thermodynamic properties (temperature, pressure, density and entropy) follow the expectations for a typical cool core cluster (Section \ref{sec:an-spec}). The global properties of the cluster are $kT=4.31 \pm 0.05$ keV, $Z=0.60\pm0.03$ $Z_{\odot}$, $F(0.5-7\text{ keV})=1.07\pm0.01\cdot10^{-11}$ erg s$^{-1}$ cm$^{-2}$, $L(0.5-7\text{ keV})=1.66\pm0.01\cdot10^{44}$ erg s$^{-1}$. We also investigated the cooling efficiency of A2495, finding that the ratio between the cooling time and the free fall time supports a condensation of the ICM into warm gas within the 25 kpc from the X-ray peak (Section \ref{sec:coolprop}).\\
    \item We detected two pairs of X-ray cavities nearby the BCG at $\sim3\sigma$ confidence level, which are likely the results of subsequent AGN outbursts. We calculated the ages and the mechanical powers of the cavities (Section \ref{sec:cav-prop}), concluding that they are able to counterbalance the radiative losses of the ICM ($P_{cav} = 4.7\pm1.3\times10^{43}$~erg~s$^{-1}$; $L_{cool} = 5.7\pm0.1\times10^{43}$~erg~s$^{-1}$) (Section \ref{sez:heat-cool}). This indicates that on the long term the feedback cycle is still efficient in A2495, despite the transitory offsets between the central engine of the BCG and the gas fuelling the AGN. 
    \item In order to probe the dynamical state of the ICM we investigated the presence of surface brightness edges and temperature discontinuities in the $Chandra$ image. We found a significant density jump ($jump=1.14\pm0.02$) in the SB profile, located at 58$^{+0.9}_{-4.2}$ kpc east from the center. The spectral analysis across this region shows a temperature and pressure configuration that resembles that of a cold front discontinuity (Section \ref{sez:cf}). Moreover, the spectral maps of temperature and entropy (Section \ref{sez:specmap}) further support this conclusion.
    \item We exclude that the offsets are caused by a mass uplift of warm/hot central gas by the cavity uprise in the cluster atmosphere, because (a) the cavities could displace at maximum a gas mass that is three times smaller than that of the offset ICM peak, and because (b) the morphology and relative position of the ICM peak, of the H$\alpha$ peak, and of the X-ray cavities is not consistent with an uplift scenario. Rather, the new data are in agreement with a {\it sloshing regulated feedback cycle}, as previously proposed by \citet{pasini_19}. Sloshing could momentarily move the ICM and warm gas peaks away from the BCG, creating the present triple offset configuration. Past passages of the gas peaks onto the BCG would have periodically fuelled the central AGN. In this respect, we determined that the sloshing timescales ($t_{sloshing}=61.1\pm 15.8\,(\pm43.2) \text{ Myr}$) is comparable with the outburst interval between the two cavity pairs ($\Delta t_{cav}=39.8\pm10.7\,(\pm19.8) \text{ Myr}$). This does not contradict the idea that the AGN feedback timescales may reflect the periodic sloshing of the gas on the BCG.
\end{itemize}


\begin{acknowledgments}
We thank the reviewer for their useful suggestions, which improved this work. Support for this work was provided by the National Aeronautics and Space Administration through Chandra Award Number GO1-22125X issued by the Chandra X-ray Center, which is operated by the Smithsonian Astrophysical Observatory for and on behalf of the National Aeronautics Space Administration under contract NAS8-03060. This paper employs a list of Chandra datasets, obtained by the Chandra X-ray Observatory, contained in~\dataset[DOI: 10.25574]{https://doi.org/10.25574/cdc.176}. M. Gaspari acknowledges partial support by NASA HST GO-15890.020/023-A, and the \textit{BlackHoleWeather} program. 
\end{acknowledgments}

\newpage

\appendix

\section{Cavity Detection}
This appendix presents the analysis performed to detect the X-ray cavities. The peculiar and asymmetric morphology of the ICM makes a reliable estimation of the surface brightness deficit and its significance not trivial, even with the high number of counts provided by the new $Chandra$ observations. We adopted three methods, the first based on the azimuthal comparison shown in \citet{larondo_15, macconi_22}, the second exploiting only the local value of the ICM around the putative cavity (as done in \citet{ubertosi2021b}), and the last performing a radial comparison. In all cases, the number of counts from a region containing the depression and from a region without them were used to calculate the significance with the following definition:

\begin{equation}
    \sigma_{cav}=\frac{|C_d-C_l|}{\sqrt{C_d+C_l}}
\end{equation}

where $C_d$ and $C_l$ are respectively the counts from the depression region and the local ICM emission.

\begin{figure}
    \centering
    \includegraphics[scale=0.4]{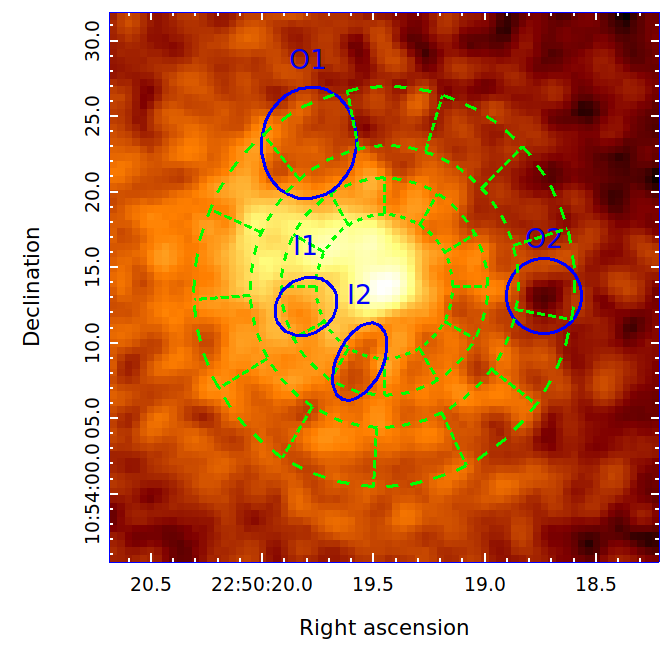}
    \caption{0.5-2 keV image with the elliptical sectors used for the azimuthal analysis of the surface brightness depressions, for both the internal and external cavities.}
    \label{fig:ep-az}
\end{figure}

\paragraph{AZIMUTHAL COMPARISON}
We extracted the counts from elliptical sectors (see Figure \ref{fig:ep-az}). Then, the counts in the region of the putative depression are compared with the average azimuthal value at the same distance from the X-ray peak. 
This is the method used by \citet{larondo_15, macconi_22} but the irregular morphology of the ICM made this analysis not trivial. We obtained the following results: O1 2.8$\sigma$, O2 5.6$\sigma$, I1 0.6$\sigma$, I2 2.2$\sigma$. 
For example, the cavity I1 likely suffers from projection effects of the surrounding bright gas, and the eastern side of the cluster is systematically brighter than the region at the same distance from the center westward.

\begin{figure}
    \centering
    \subfigure[]{
    \includegraphics[scale=0.5]{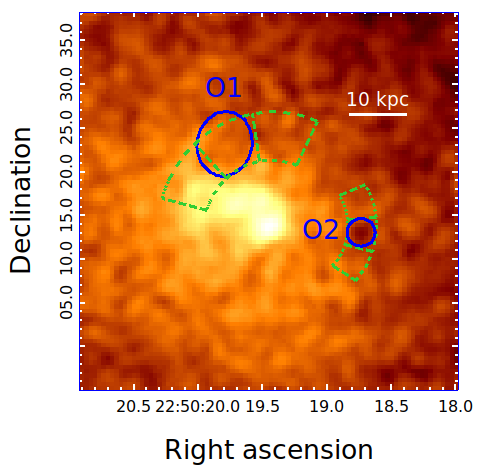}}
    \subfigure[]{
    \includegraphics[scale=0.5]{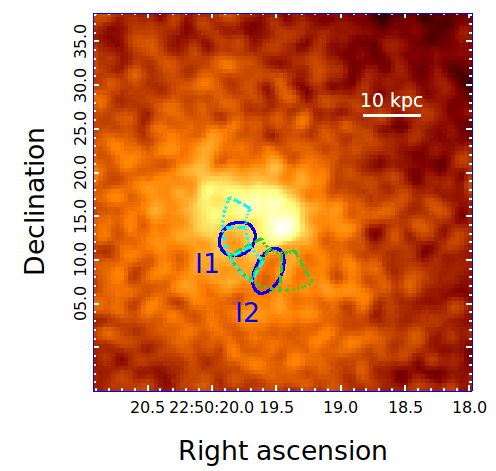}}
    \caption{0.5-2 keV $Chandra$ counts images with the regions used for the comparison between the depression and the value of the local ICM around it.}
    \label{fig:ep-ad}
\end{figure}

\paragraph{LOCAL COMPARISON}
For this method we consider only the regions next to the depression, in order to reduce the biases caused from the azimuthal irregularity of the ICM.
In Figure \ref{fig:ep-ad} are shown the region used for each cavity. The significance obtained with this method are reported in Table \ref{tab:cav}, and are: O1 3.8$\sigma$, O2 2.9$\sigma$, I1 3.3$\sigma$, I2 2.5$\sigma$.

\begin{figure}[ht!]
    \centering
    \subfigure[]{
    \includegraphics[scale=0.35]{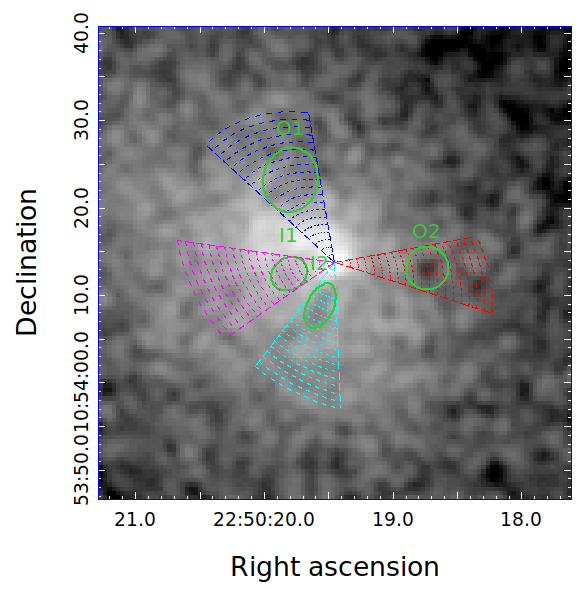}}
    \subfigure[]{
    \includegraphics[scale=0.52]{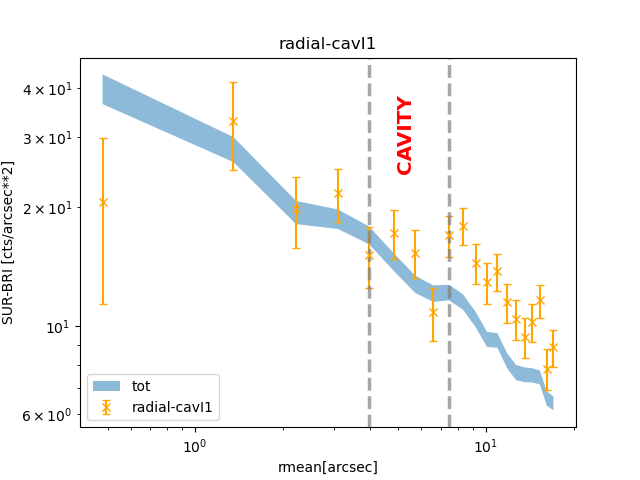}}
    \subfigure[]{
    \includegraphics[scale=0.52]{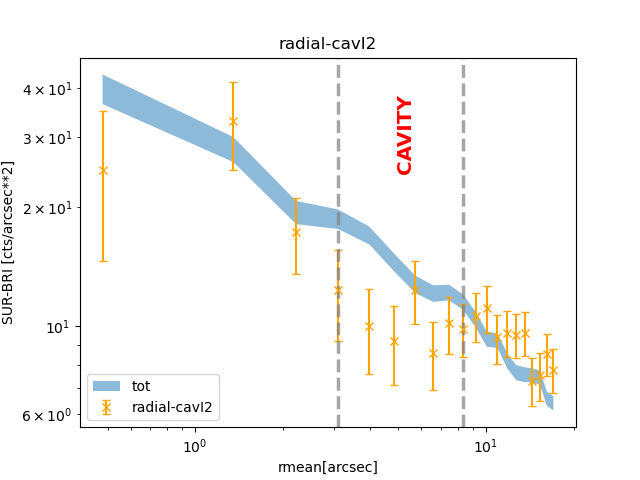}}
    \subfigure[]{
    \includegraphics[scale=0.52]{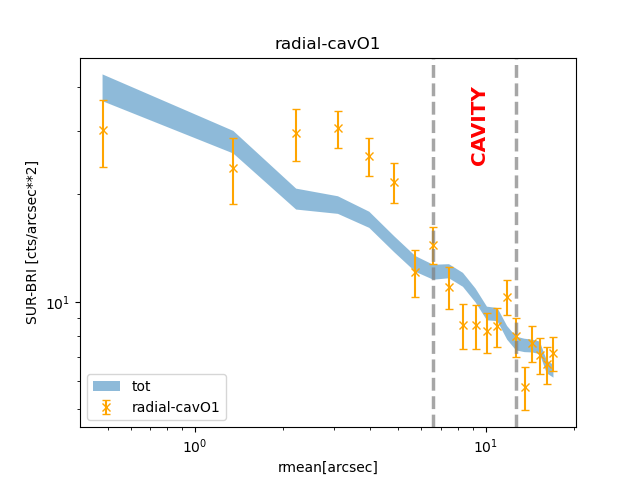}}
    \subfigure[]{
    \includegraphics[scale=0.52]{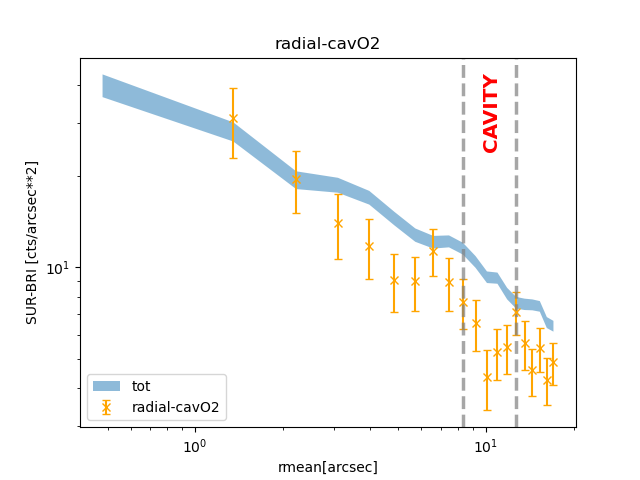}}
    \caption{(a) 0.5-2 keV image with the sectors used for the radial comparison. (b-c-d-e) Radial profiles of the sector enclosing the depression (orange crosses) against the total radial profile (blue shaded).}
    \label{fig:radial-plot}
\end{figure}

\paragraph{RADIAL COMPARISON}
At last, we tried to locate the depression in the radial direction similarly done in \citet{doria_12}, Figure 10. Surface brightness profiles were extracted using a 1-arcsec bin width. With this resolution it is possible to enclose the cavity region with more than one bin. The profile along the cavity direction is then compared with the azimuthally averaged total profile (see Figure \ref{fig:radial-plot}). The significance values obtained with this analysis are the following: O1 0.3$\sigma$, O2 3.3$\sigma$, I1 -0.6$\sigma$, I2 1.6$\sigma$.

\clearpage

\section{Projected analysis: Best-fit parameters}

\begin{longtable}{|c|c|c|c|c|c|}
\hline
Annulus & R [kpc] & kT [keV] & Z [$Z_{\odot}$] & counts & $\chi^2/d.o.f.$ \\
\hline
\rule[-2mm]{0mm}{3mm}
 1 & 12 & 2.4$_{-0.1}^{+0.1}$ & 0.9$_{-0.2}^{+0.2}$ & 3400 & 1.24 \\
\hline
\rule[-2mm]{0mm}{3mm}
 2 & 19 & 3.5$_{-0.2}^{+0.2}$ & 0.9$_{-0.2}^{+0.3}$ & 3600 & 0.83 \\
\hline
\rule[-2mm]{0mm}{3mm}
 3 & 25 & 3.9$_{-0.2}^{+0.2}$ & 1.2$_{-0.3}^{+0.4}$ & 3713 & 0.91 \\
\hline
\rule[-2mm]{0mm}{3mm}
 4 & 31 & 3.6$_{-0.2}^{+0.2}$ & 0.5$_{-0.2}^{+0.2}$ & 3492 & 0.99 \\
\hline
\rule[-2mm]{0mm}{3mm}
 5 & 36 & 3.7$_{-0.2}^{+0.2}$ & 0.3$_{-0.2}^{+0.2}$ & 3575 & 0.98 \\
\hline
\rule[-2mm]{0mm}{3mm}
 6 & 41 & 3.9$_{-0.2}^{+0.2}$ & 0.6$_{-0.2}^{+0.3}$ & 3586 & 0.97 \\
\hline
\rule[-2mm]{0mm}{3mm}
 7 & 46 & 4.1$_{-0.2}^{+0.2}$ & 0.7$_{-0.2}^{+0.3}$ & 3487 & 0.88 \\
\hline
\rule[-2mm]{0mm}{3mm}
 8 & 52 & 4.3$_{-0.2}^{+0.2}$ & 0.7$_{-0.2}^{+0.3}$ & 3482 & 0.88 \\
\hline
\rule[-2mm]{0mm}{3mm}
 9 & 57 & 4.4$_{-0.3}^{+0.3}$ & 0.6$_{-0.3}^{+0.3}$ & 3355 & 1.05 \\
\hline
\rule[-2mm]{0mm}{3mm}
 10 & 63 & 4.3$_{-0.3}^{+0.3}$ & 0.8$_{-0.3}^{+0.4}$ & 3508 & 0.95 \\
\hline
\rule[-2mm]{0mm}{3mm}
 11 & 70 & 4.2$_{-0.2}^{+0.2}$ & 0.5$_{-0.2}^{+0.2}$ & 3843 & 0.84 \\
\hline
\rule[-2mm]{0mm}{3mm}
 12 & 76 & 4.4$_{-0.3}^{+0.3}$ & 0.5$_{-0.2}^{+0.3}$ & 3574 & 0.88 \\
\hline
\rule[-2mm]{0mm}{3mm}
 13 & 83 & 4.8$_{-0.3}^{+0.3}$ & 0.9$_{-0.3}^{+0.3}$ & 3427 & 1.03 \\
\hline
\rule[-2mm]{0mm}{3mm}
 14 & 90 & 5.0$_{-0.3}^{+0.3}$ & 0.7$_{-0.3}^{+0.3}$ & 3526 & 0.97 \\
\hline
\rule[-2mm]{0mm}{3mm}
 15 & 98 & 5.5$_{-0.3}^{+0.4}$ & 1.2$_{-0.3}^{+0.4}$ & 3465 & 0.83 \\
\hline
\rule[-2mm]{0mm}{3mm}
 16 & 106 & 4.6$_{-0.3}^{+0.4}$ & 0.1$_{-0.1}^{+0.2}$ & 3503 & 1.31 \\
\hline
\rule[-2mm]{0mm}{3mm}
 17 & 114 & 4.3$_{-0.3}^{+0.3}$ & 0.7$_{-0.3}^{+0.3}$ & 3549 & 1.06 \\
\hline
\rule[-2mm]{0mm}{3mm}
 18 & 123 & 5.1$_{-0.3}^{+0.4}$ & 2.0$_{-0.6}^{+0.8}$ & 3311 & 1.02 \\
\hline
\rule[-2mm]{0mm}{3mm}
 19 & 132 & 4.7$_{-0.3}^{+0.3}$ & 0.5$_{-0.2}^{+0.3}$ & 3411 & 0.86 \\
\hline
\rule[-2mm]{0mm}{3mm}
 20 & 143 & 4.8$_{-0.3}^{+0.4}$ & 0.2$_{-0.2}^{+0.2}$ & 3737 & 0.95 \\
\hline
\rule[-2mm]{0mm}{3mm}
 21 & 157 & 4.4$_{-0.3}^{+0.3}$ & 0.5$_{-0.2}^{+0.3}$ & 3884 & 1.05 \\
\hline
\rule[-2mm]{0mm}{3mm}
 22 & 173 & 5.1$_{-0.4}^{+0.4}$ & 0.5$_{-0.3}^{+0.3}$ & 3891 & 0.93 \\
\hline
\rule[-2mm]{0mm}{3mm}
 23 & 188 & 4.7$_{-0.4}^{+0.4}$ & 0.4$_{-0.2}^{+0.3}$ & 3451 & 1.18 \\
\hline
\rule[-2mm]{0mm}{3mm}
 24 & 207 & 4.9$_{-0.4}^{+0.4}$ & 0.2$_{-0.2}^{+0.3}$ & 3817 & 1.04 \\
\hline
\rule[-2mm]{0mm}{3mm}
 25 & 226 & 4.6$_{-0.4}^{+0.4}$ & 0.4$_{-0.3}^{+0.3}$ & 3453 & 1.16 \\
\hline
\rule[-2mm]{0mm}{3mm}
 26 & 249 & 5.1$_{-0.4}^{+0.5}$ & 0.5$_{-0.3}^{+0.3}$ & 4035 & 1.18 \\
\hline
\rule[-2mm]{0mm}{3mm}
 27 & 276 & 5.7$_{-0.5}^{+0.6}$ & 0.4$_{-0.3}^{+0.4}$ & 3859 & 1.04 \\
\hline
\rule[-2mm]{0mm}{3mm}
 28 & 304 & 5.1$_{-0.5}^{+0.6}$ & 0.0$_{-0.0}^{+0.3}$ & 3516 & 1.06 \\
\hline
\rule[-2mm]{0mm}{3mm}
 29 & 336 & 5.0$_{-0.5}^{+0.6}$ & 0.3$_{-0.3}^{+0.3}$ & 3597 & 0.98 \\
\hline
\rule[-2mm]{0mm}{3mm}
 30 & 372 & 5.8$_{-0.7}^{+0.8}$ & 0.4$_{-0.3}^{+0.4}$ & 3580 & 1.05 \\
\hline
\rule[-2mm]{0mm}{3mm}
 31 & 429 & 6.8$_{-0.8}^{+1.0}$ & 0.6$_{-0.4}^{+0.5}$ & 4330 & 0.93 \\
\hline
\caption{Best-fit parameter for the projected spectral analysis. (1) Annulus number; (2) mean distance of each annulus from the X-ray peak; (3) temperature in keV; (4) metallicity in solar unit ($Z_{\odot}$); (5) net counts; fit goodness ($\chi^2/d.o.f.$).}
\label{tab:projected}
\end{longtable}

\newpage

\section{Deprojected analysis: Best-fit parameters}

\begin{longtable}{|c|c|c|c|c|c|c|c|}
\hline
Region & R & kT & $n_e$ & P & K & counts & $\chi^2/d.o.f.$ \\
\hline
\rule[-2mm]{0mm}{3mm}
 1 & 7 & 1.3$_{-0.1}^{+0.2}$ & 6.9$_{-1.5}^{+1.5}$ & 1.6$_{-0.4}^{+0.4}$ & 7.5$_{-1.4}^{+1.5}$& 1332 & 1.13 \\
\hline
\rule[-2mm]{0mm}{3mm}
 2 & 13 & 2.2$_{-0.3}^{+0.4}$ & 2.8$_{-0.6}^{+0.6}$ & 1.1$_{-0.3}^{+0.3}$ & 23.8$_{-5.0}^{+6.0}$& 2431 & 0.78 \\
\hline
\rule[-2mm]{0mm}{3mm}
 3 & 19 & 2.9$_{-0.6}^{+1.2}$ & 1.9$_{-0.2}^{+0.2}$ & 1.0$_{-0.3}^{+0.4}$ & 41.2$_{-9.5}^{+16.7}$& 2915 & 0.98 \\
\hline
\rule[-2mm]{0mm}{3mm}
 4 & 25 & 3.8$_{-0.7}^{+0.8}$ & 2.2$_{-0.2}^{+0.2}$ & 1.5$_{-0.3}^{+0.4}$ & 49.5$_{-9.3}^{+11.2}$& 3713 & 0.92 \\
\hline
\rule[-2mm]{0mm}{3mm}
 5 & 30 & 3.0$_{-0.4}^{+0.6}$ & 1.8$_{-0.2}^{+0.2}$ & 1.0$_{-0.2}^{+0.2}$ & 43.4$_{-7.1}^{+9.7}$& 3492 & 0.97 \\
\hline
\rule[-2mm]{0mm}{3mm}
 6 & 36 & 3.6$_{-0.6}^{+0.8}$ & 1.5$_{-0.2}^{+0.2}$ & 1.0$_{-0.2}^{+0.3}$ & 58.3$_{-10.8}^{+13.8}$& 3575 & 0.97 \\
\hline
\rule[-2mm]{0mm}{3mm}
 7 & 41 & 3.3$_{-0.5}^{+0.7}$ & 1.6$_{-0.2}^{+0.2}$ & 1.0$_{-0.2}^{+0.2}$ & 51.7$_{-9.2}^{+11.8}$& 3586 & 0.95 \\
\hline
\rule[-2mm]{0mm}{3mm}
 8 & 46 & 3.6$_{-0.7}^{+0.9}$ & 1.1$_{-0.2}^{+0.2}$ & 0.76$_{-0.20}^{+0.24}$ & 71.6$_{-16.3}^{+19.6}$& 3487 & 0.86 \\
\hline
\rule[-2mm]{0mm}{3mm}
 9 & 52 & 3.9$_{-0.7}^{+1.0}$ & 1.1$_{-0.2}^{+0.2}$ & 0.79$_{-0.19}^{+0.23}$ & 80.0$_{-16.5}^{+21.7}$& 3482 & 0.86 \\
\hline
\rule[-2mm]{0mm}{3mm}
 10 & 57 & 4.7$_{-1.0}^{+1.4}$ & 1.0$_{-0.1}^{+0.1}$ & 0.87$_{-0.22}^{+0.28}$ & 99.3$_{-23.2}^{+31.3}$& 3355 & 1.02 \\
\hline
\rule[-2mm]{0mm}{3mm}
 11 & 63 & 4.0$_{-0.7}^{+1.2}$ & 0.93$_{-0.10}^{+0.10}$ & 0.67$_{-0.15}^{+0.21}$ & 89.4$_{-17.8}^{+27.1}$& 3508 & 0.93 \\
\hline
\rule[-2mm]{0mm}{3mm}
 12 & 70 & 3.8$_{-0.7}^{+1.0}$ & 0.81$_{-0.12}^{+0.12}$ & 0.57$_{-0.13}^{+0.17}$ & 94.7$_{-19.4}^{+26.2}$& 3843 & 0.83 \\
\hline
\rule[-2mm]{0mm}{3mm}
 13 & 76 & 3.5$_{-0.6}^{+1.0}$ & 0.77$_{-0.11}^{+0.11}$ & 0.50$_{-0.11}^{+0.16}$ & 89.9$_{-17.4}^{+28.1}$& 3574 & 0.86 \\
\hline
\rule[-2mm]{0mm}{3mm}
 14 & 83 & 4.5$_{-1.1}^{+1.4}$ & 0.64$_{-0.07}^{+0.07}$ & 0.53$_{-0.14}^{+0.18}$ & 129.0$_{-32.4}^{+42.6}$& 3427 & 1.01 \\
\hline
\rule[-2mm]{0mm}{3mm}
 15 & 90 & 4.2$_{-0.9}^{+1.7}$ & 0.55$_{-0.06}^{+0.06}$ & 0.42$_{-0.10}^{+0.18}$ & 133.0$_{-30.2}^{+56.1}$& 3526 & 0.95 \\
\hline
\rule[-2mm]{0mm}{3mm}
 16 & 98 & 6.7$_{-1.7}^{+2.3}$ & 0.57$_{-0.06}^{+0.06}$ & 0.70$_{-0.19}^{+0.25}$ & 209.0$_{-54.0}^{+73.9}$& 3465 & 0.84 \\
\hline
\rule[-2mm]{0mm}{3mm}
 17 & 106 & 6.5$_{-1.5}^{+2.5}$ & 0.45$_{-0.05}^{+0.05}$ & 0.53$_{-0.14}^{+0.21}$ & 237.0$_{-58.9}^{+91.8}$& 3503 & 1.31 \\
\hline
\rule[-2mm]{0mm}{3mm}
 18 & 114 & 4.1$_{-0.6}^{+0.9}$ & 0.45$_{-0.05}^{+0.05}$ & 0.33$_{-0.06}^{+0.08}$ & 149.0$_{-26.0}^{+34.4}$& 3549 & 1.04 \\
\hline
\rule[-2mm]{0mm}{3mm}
 19 & 123 & 4.2$_{-0.8}^{+1.7}$ & 0.38$_{-0.04}^{+0.04}$ & 0.30$_{-0.07}^{+0.12}$ & 174.0$_{-36.8}^{+69.6}$& 3311 & 1.06 \\
\hline
\rule[-2mm]{0mm}{3mm}
 20 & 132 & 4.1$_{-0.7}^{+0.7}$ & 0.39$_{-0.04}^{+0.04}$ & 0.29$_{-0.06}^{+0.06}$ & 166.0$_{-30.5}^{+30.6}$& 3411 & 0.84 \\
\hline
\rule[-2mm]{0mm}{3mm}
 21 & 143 & 6.0$_{-1.3}^{+1.7}$ & 0.32$_{-0.03}^{+0.03}$ & 0.36$_{-0.08}^{+0.11}$ & 277.0$_{-60.7}^{+78.4}$& 3737 & 0.95 \\
\hline
\rule[-2mm]{0mm}{3mm}
 22 & 157 & 3.9$_{-0.6}^{+0.9}$ & 0.32$_{-0.04}^{+0.04}$ & 0.23$_{-0.04}^{+0.06}$ & 179.0$_{-30.1}^{+42.5}$& 3884 & 1.03 \\
\hline
\rule[-2mm]{0mm}{3mm}
 23 & 173 & 4.9$_{-1.0}^{+1.2}$ & 0.26$_{-0.03}^{+0.03}$ & 0.23$_{-0.05}^{+0.06}$ & 264.0$_{-56.9}^{+67.9}$& 3849 & 0.92 \\
\hline
\rule[-2mm]{0mm}{3mm}
 24 & 207 & 5.1$_{-0.7}^{+0.8}$ & 0.19$_{-0.02}^{+0.02}$ & 0.18$_{-0.03}^{+0.03}$ & 329.0$_{-49.8}^{+55.4}$& 6727 & 1.15 \\
\hline
\rule[-2mm]{0mm}{3mm}
 25 & 249 & 4.2$_{-0.5}^{+0.6}$ & 0.14$_{-0.02}^{+0.02}$ & 0.11$_{-0.02}^{+0.02}$ & 335.0$_{-45.0}^{+54.1}$& 6377 & 1.21 \\
\hline
\rule[-2mm]{0mm}{3mm}
 26 & 304 & 5.6$_{-0.8}^{+0.9}$ & 0.09$_{-0.01}^{+0.01}$ & 0.09$_{-0.02}^{+0.02}$ & 586.0$_{-89.8}^{+101.0}$& 6155 & 1.24 \\
\hline
\rule[-2mm]{0mm}{3mm}
 27 & 429 & 5.8$_{-0.3}^{+0.4}$ & 0.08$_{-0.01}^{+0.01}$ & 0.08$_{-0.01}^{+0.01}$ & 670.0$_{-60.5}^{+62.0}$& 9508 & 0.91 \\
\hline
\caption{Best-fit parameter for the deprojected spectral analysis. (1) Annulus number; (2) mean distance of each annulus from the X-ray peak in kpc; (3) temperature [keV]; (4) electron density $[10^{-2}$ cm$^{-3}]$; (5) Gas pressure $[10^{-1}$ keV cm$^{-3}]$; (6) Gas entropy [keV cm$^{2}$] (5) net counts; fit goodness ($\chi^2/d.o.f.$).}
\label{tab:deprojected}
\end{longtable}

\section{Testing the Hidden Cooling Flow model in A2495}\label{app:hcf}
\begin{figure}[h!]
  \centering
  \includegraphics[width=0.5\linewidth]{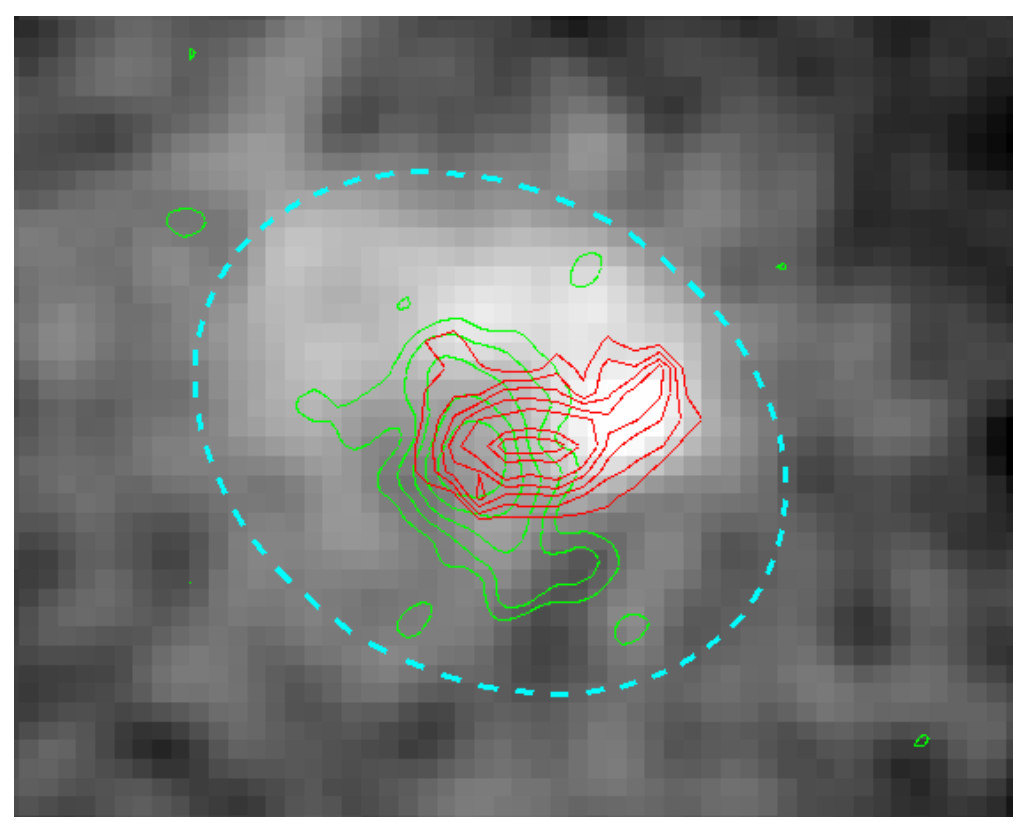}
  \caption{{\it Chandra} 0.5 -- 7 keV image centered on the core of A2495. 5~GHz contours from \citet{pasini_19} and H$\alpha$ total intensity contours from \citet{hamer_16} are overlaid in green and red, respectively. The dashed-cyan ellipse shows the spectral extraction region used to test the Hidden Cooling Flow model \citep{fabian2022}. This region encompasses the X-ray peak, the bright arc-shaped structure, and the dust lane seen in HST data. The major semi-axis of this region is $\sim$20~kpc.}\label{fig:hiddenCF}
\end{figure}

In this Appendix we present the details of the spectral fit to the central X-ray emission in A2495 using the Hidden Cooling Flow model \citep{fabian2022,fabian2023}. As reported in Sect. \ref{sec:coolprop}, we extracted the spectrum of the innermost 20~kpc, using a region that encompasses the X-ray peak, the X-ray bright arc-shaped structure, the central AGN, and the dust lane. This region is shown in Fig. \ref{fig:hiddenCF}. We fitted the spectrum of this region in \texttt{Xspec} using the model described in Fabian et al. (2022), that is \texttt{tbabs$\ast$(apec + mkcflow$_{u}$ + mlayer$\ast$mkcflow$_{a}$)}. For completeness, we also tested an alternative model, \texttt{tbabs$\ast$(apec + ztbabs$\ast$mkcflow)}. Results of these tests are reported in Tab. \ref{tab:hiddenCF}. 
\begin{table}
    \centering
    \begin{tabular}{|c|c|}
    \hline
        \textbf{\citet{fabian2022}} & \textbf{Our model + absorption}\\
         \texttt{tbabs*(apec + mkcflow + mlayer*mkcflow)} & \texttt{tbabs*(apec + ztbabs*mkcflow)} \\
         \hline
         
         nH(\texttt{tbabs}): 4.41$\cdot10^{20}$f & nH(\texttt{tbabs}): 4.41$\cdot10^{20}$f\\
         kT(\texttt{apec}): $3.32_{-0.30}^{+0.34}$ keV& kT(\texttt{apec}): $3.13_{-0.29}^{+0.52}$ keV\\
         Z(\texttt{apec}): $1.46_{-0.31}^{+0.38}$ Z$\odot$&  Z(\texttt{apec}): $0.87_{-0.26}^{+0.29}$ Z $\odot$\\
         Redshift(\texttt{apec}): 0.07923f &  Redshift(\texttt{apec}): 0.07923f\\
         norm(\texttt{apec}): $6.99_{-6.00}^{+4.23}\cdot10^{-5}$ cm$^{-3}$& norm(\texttt{apec}): $1.04_{-0.83}^{+0.57}\cdot10^{-4}$ cm$^{-3}$\\
         LowT(\texttt{mkcflow}): 1.0f keV & nH(\texttt{ztbabs}): $0.39_{-0.07}^{+0.08}\cdot10^{22}$ cm$^{-3}$\\
         HighT(\texttt{mkcflow}): = kT(\texttt{apec}) & Redshift(\texttt{ztbabs}): = Redshift(\texttt{apec})\\
         Z(\texttt{mkcflow}): = Z(\texttt{apec})& LowT(\texttt{mkcflow}): 0.0808f keV\\
         Redshift(\texttt{mkcflow}): = Redshift(\texttt{apec}) & HighT(\texttt{mkcflow}): = kT(\texttt{apec})\\
         Norm(\texttt{mkcflow}): $<$ 6.0 M$_{\odot}$/yr & Z(\texttt{mkcflow}): = Z(\texttt{apec})\\
         nH(\texttt{mlayer}): $0.79_{-0.10}^{+0.20}\cdot10^{22}$ cm$^{-3}$ & Redshift(\texttt{mkcflow}): 0.07923f\\
         Redshift(\texttt{mlayer}): = Redshift(\texttt{apec}) & Norm(\texttt{mkcflow}): $11.11_{-3.15}^{+2.80}$ M$_\odot$/yr\\
         LowT(\texttt{mkcflow}): 0.1f keV & \\
         HighT(\texttt{mkcflow}): = kT(\texttt{apec}) & \\
         Z(\texttt{mkcflow}): =  Z(\texttt{apec})& \\
         Redshift(\texttt{mkcflow}): 0.07923f & \\
         Norm(\texttt{mkcflow}): $10.84_{-2.27}^{+2.08}$ M$_{\odot}$/yr& \\
         \hline
         $\chi^{2} /d.o.f = 177.06/138 = 1.29$& $\chi^{2} /d.o.f = 182.67/139 = 1.32$ \\
         \hline
    \end{tabular}
    \caption{Spectral fit to the X-ray emission within the ellipse shown in Fig. \ref{fig:hiddenCF}.}
    \label{tab:hiddenCF}
\end{table}
\vspace{5mm}
\facilities{{\it Chandra.}}


\software{
       {\bf XSPEC \citep{arnaud_96},
       CIAO \citep{fruscione_06},
       PROFFIT \citep{eckert_2011}.}
       % astropy \citep{2013A&A...558A..33A,2018AJ....156..123A},  
        %Cloudy \citep{2013RMxAA..49..137F},      
         }





\bibliography{A2495}{}

\begin{thebibliography}{}
\expandafter\ifx\csname natexlab\endcsname\relax\def\natexlab#1{#1}\fi
\providecommand{\url}[1]{\href{#1}{#1}}
\providecommand{\dodoi}[1]{doi:~\href{http://doi.org/#1}{\nolinkurl{#1}}}
\providecommand{\doeprint}[1]{\href{http://ascl.net/#1}{\nolinkurl{http://ascl.net/#1}}}
\providecommand{\doarXiv}[1]{\href{https://arxiv.org/abs/#1}{\nolinkurl{https://arxiv.org/abs/#1}}}

\bibitem[{{Akritas} \& {Bershady}(1996)}]{akritas_96}
{Akritas}, M.~G., \& {Bershady}, M.~A. 1996, \apj, 470, 706,
  \dodoi{10.1086/177901}

\bibitem[{{Arnaud}(1996)}]{arnaud_96}
{Arnaud}, K.~A. 1996, in Astronomical Society of the Pacific Conference Series,
  Vol. 101, Astronomical Data Analysis Software and Systems V, ed. G.~H.
  {Jacoby} \& J.~{Barnes}, 17

\bibitem[{{Asplund} {et~al.}(2009){Asplund}, {Grevesse}, {Sauval}, \&
  {Scott}}]{asplund_09}
{Asplund}, M., {Grevesse}, N., {Sauval}, A.~J., \& {Scott}, P. 2009, \araa, 47,
  481, \dodoi{10.1146/annurev.astro.46.060407.145222}

\bibitem[{{Biava} {et~al.}(2021){Biava}, {Brienza}, {Bonafede}, {Gitti},
  {Bonnassieux}, {Harwood}, {Edge}, {Riseley}, \& {Vantyghem}}]{biava_21}
{Biava}, N., {Brienza}, M., {Bonafede}, A., {et~al.} 2021, \aap, 650, A170,
  \dodoi{10.1051/0004-6361/202040063}

\bibitem[{{B\^irzan} {et~al.}(2017){B\^irzan}, {Rafferty}, {Br{\"u}ggen}, \&
  {Intema}}]{birzan_17}
{B\^irzan}, L., {Rafferty}, D.~A., {Br{\"u}ggen}, M., \& {Intema}, H.~T. 2017,
  \mnras, 471, 1766, \dodoi{10.1093/mnras/stx1505}

\bibitem[{{B\^irzan} {et~al.}(2004){B\^irzan}, {Rafferty}, {McNamara}, {Wise},
  \& {Nulsen}}]{birzan_04}
{B\^irzan}, L., {Rafferty}, D.~A., {McNamara}, B.~R., {Wise}, M.~W., \&
  {Nulsen}, P.~E.~J. 2004, \apj, 607, 800, \dodoi{10.1086/383519}

\bibitem[{{B{\^\i}rzan} {et~al.}(2012){B{\^\i}rzan}, {Rafferty}, {Nulsen},
  {McNamara}, {R{\"o}ttgering}, {Wise}, \& {Mittal}}]{birzan_12}
{B{\^\i}rzan}, L., {Rafferty}, D.~A., {Nulsen}, P.~E.~J., {et~al.} 2012,
  \mnras, 427, 3468, \dodoi{10.1111/j.1365-2966.2012.22083.x}

\bibitem[{{Bruno} {et~al.}(2019){Bruno}, {Gitti}, {Zanichelli}, \&
  {Gregorini}}]{bruno_19}
{Bruno}, L., {Gitti}, M., {Zanichelli}, A., \& {Gregorini}, L. 2019, \aap, 631,
  A173, \dodoi{10.1051/0004-6361/201936240}

\bibitem[{{Campitiello} {et~al.}(2022){Campitiello}, {Ettori}, {Lovisari},
  {Bartalucci}, {Eckert}, {Rasia}, {Rossetti}, {Gastaldello}, {Pratt},
  {Maughan}, {Pointecouteau}, {Sereno}, {Biffi}, {Borgani}, {De Luca}, {De
  Petris}, {Gaspari}, {Ghizzardi}, {Mazzotta}, \&
  {Molendi}}]{2022A&A...665A.117C}
{Campitiello}, M.~G., {Ettori}, S., {Lovisari}, L., {et~al.} 2022, \aap, 665,
  A117, \dodoi{10.1051/0004-6361/202243470}

\bibitem[{{Crawford} {et~al.}(1999){Crawford}, {Allen}, {Ebeling}, {Edge}, \&
  {Fabian}}]{crawford_99}
{Crawford}, C.~S., {Allen}, S.~W., {Ebeling}, H., {Edge}, A.~C., \& {Fabian},
  A.~C. 1999, \mnras, 306, 857, \dodoi{10.1046/j.1365-8711.1999.02583.x}

\bibitem[{{Cui} {et~al.}(2016){Cui}, {Power}, {Biffi}, {Borgani}, {Murante},
  {Fabjan}, {Knebe}, {Lewis}, \& {Poole}}]{cui_16}
{Cui}, W., {Power}, C., {Biffi}, V., {et~al.} 2016, \mnras, 456, 2566,
  \dodoi{10.1093/mnras/stv2839}

\bibitem[{{Doria} {et~al.}(2012){Doria}, {Gitti}, {Ettori}, {Brighenti},
  {Nulsen}, \& {McNamara}}]{doria_12}
{Doria}, A., {Gitti}, M., {Ettori}, S., {et~al.} 2012, \apj, 753, 47,
  \dodoi{10.1088/0004-637X/753/1/47}

\bibitem[{{Ebeling} {et~al.}(1998){Ebeling}, {Edge}, {Bohringer}, {Allen},
  {Crawford}, {Fabian}, {Voges}, \& {Huchra}}]{ebeling_98}
{Ebeling}, H., {Edge}, A.~C., {Bohringer}, H., {et~al.} 1998, \mnras, 301, 881,
  \dodoi{10.1046/j.1365-8711.1998.01949.x}

\bibitem[{{Eckert} {et~al.}(2016){Eckert}, {Jauzac}, {Vazza}, {Owers}, {Kneib},
  {Tchernin}, {Intema}, \& {Knowles}}]{eckert_16}
{Eckert}, D., {Jauzac}, M., {Vazza}, F., {et~al.} 2016, \mnras, 461, 1302,
  \dodoi{10.1093/mnras/stw1435}

\bibitem[{{Eckert} {et~al.}(2011){Eckert}, {Molendi}, \&
  {Paltani}}]{eckert_2011}
{Eckert}, D., {Molendi}, S., \& {Paltani}, S. 2011, \aap, 526, A79,
  \dodoi{10.1051/0004-6361/201015856}

\bibitem[{Eilek(2014)}]{eilek_14}
Eilek, J.~A. 2014, New Journal of Physics, 16, 045001,
  \dodoi{10.1088/1367-2630/16/4/045001}

\bibitem[{{Ettori} {et~al.}(2013){Ettori}, {Gastaldello}, {Gitti},
  {O'Sullivan}, {Gaspari}, {Brighenti}, {David}, \& {Edge}}]{ettori_13}
{Ettori}, S., {Gastaldello}, F., {Gitti}, M., {et~al.} 2013, \aap, 555, A93,
  \dodoi{10.1051/0004-6361/201321107}

\bibitem[{{Fabian}(1994)}]{fabian_94a}
{Fabian}, A.~C. 1994, \araa, 32, 277,
  \dodoi{10.1146/annurev.aa.32.090194.001425}

\bibitem[{{Fabian} {et~al.}(2022){Fabian}, {Ferland}, {Sanders}, {McNamara},
  {Pinto}, \& {Walker}}]{fabian2022}
{Fabian}, A.~C., {Ferland}, G.~J., {Sanders}, J.~S., {et~al.} 2022, \mnras,
  515, 3336, \dodoi{10.1093/mnras/stac2003}

\bibitem[{{Fabian} {et~al.}(2023){Fabian}, {Sanders}, {Ferland}, {McNamara},
  {Pinto}, \& {Walker}}]{fabian2023}
{Fabian}, A.~C., {Sanders}, J.~S., {Ferland}, G.~J., {et~al.} 2023, \mnras,
  521, 1794, \dodoi{10.1093/mnras/stad507}

\bibitem[{Fruscione {et~al.}(2006)Fruscione, McDowell, Allen, Brickhouse,
  Burke, Davis, Durham, Elvis, Galle, Harris, Huenemoerder, Houck, Ishibashi,
  Karovska, Nicastro, Noble, Nowak, Primini, Siemiginowska, Smith, \&
  Wise}]{fruscione_06}
Fruscione, A., McDowell, J.~C., Allen, G.~E., {et~al.} 2006, in Observatory
  Operations: Strategies, Processes, and Systems, ed. D.~R. Silva \& R.~E.
  Doxsey, Vol. 6270, International Society for Optics and Photonics (SPIE), 586
  -- 597, \dodoi{10.1117/12.671760}

\bibitem[{{Gaia Collaboration}(2018)}]{gaia_18}
{Gaia Collaboration}. 2018, \aap, 616, A1, \dodoi{10.1051/0004-6361/201833051}

\bibitem[{{Gaspari} {et~al.}(2013){Gaspari}, {Ruszkowski}, \&
  {Oh}}]{gaspari_13}
{Gaspari}, M., {Ruszkowski}, M., \& {Oh}, S.~P. 2013, \mnras, 432, 3401,
  \dodoi{10.1093/mnras/stt692}

\bibitem[{{Gaspari} {et~al.}(2012){Gaspari}, {Ruszkowski}, \&
  {Sharma}}]{gaspari_12}
{Gaspari}, M., {Ruszkowski}, M., \& {Sharma}, P. 2012, \apj, 746, 94,
  \dodoi{10.1088/0004-637X/746/1/94}

\bibitem[{{Gaspari} {et~al.}(2020){Gaspari}, {Tombesi}, \&
  {Cappi}}]{gaspari_20}
{Gaspari}, M., {Tombesi}, F., \& {Cappi}, M. 2020, Nature Astronomy, 4, 10,
  \dodoi{10.1038/s41550-019-0970-1}

\bibitem[{Gaspari {et~al.}(2018)Gaspari, McDonald, Hamer, Brighenti, Temi,
  Gendron-Marsolais, Hlavacek-Larrondo, Edge, Werner, Tozzi, Sun, Stone,
  Tremblay, Hogan, Eckert, Ettori, Yu, Biffi, \& Planelles}]{gaspari_18}
Gaspari, M., McDonald, M., Hamer, S.~L., {et~al.} 2018, The Astrophysical
  Journal, 854, 167, \dodoi{10.3847/1538-4357/aaaa1b}

\bibitem[{{Ghizzardi} {et~al.}(2014){Ghizzardi}, {De Grandi}, \&
  {Molendi}}]{ghizzardi_14}
{Ghizzardi}, S., {De Grandi}, S., \& {Molendi}, S. 2014, \aap, 570, A117,
  \dodoi{10.1051/0004-6361/201424016}

\bibitem[{{Gitti} {et~al.}(2012){Gitti}, {Brighenti}, \& {McNamara}}]{gitti_12}
{Gitti}, M., {Brighenti}, F., \& {McNamara}, B.~R. 2012, Advances in Astronomy,
  2012, 950641, \dodoi{10.1155/2012/950641}

\bibitem[{Hamer {et~al.}(2012)Hamer, Edge, Swinbank, Wilman, Russell, Fabian,
  Sanders, \& Salomé}]{hamer_12}
Hamer, S.~L., Edge, A.~C., Swinbank, A.~M., {et~al.} 2012, Monthly Notices of
  the Royal Astronomical Society, 421, 3409,
  \dodoi{10.1111/j.1365-2966.2012.20566.x}

\bibitem[{{Hamer} {et~al.}(2016){Hamer}, {Edge}, {Swinbank}, {Wilman},
  {Combes}, {Salom{\'e}}, {Fabian}, {Crawford}, {Russell}, {Hlavacek-Larrondo},
  {McNamara}, \& {Bremer}}]{hamer_16}
{Hamer}, S.~L., {Edge}, A.~C., {Swinbank}, A.~M., {et~al.} 2016, \mnras, 460,
  1758, \dodoi{10.1093/mnras/stw1054}

\bibitem[{{HI4PI Collaboration} {et~al.}(2016){HI4PI Collaboration}, {Ben
  Bekhti}, {Fl{\"o}er}, {Keller}, {Kerp}, {Lenz}, {Winkel}, {Bailin},
  {Calabretta}, {Dedes}, {Ford}, {Gibson}, {Haud}, {Janowiecki}, {Kalberla},
  {Lockman}, {McClure-Griffiths}, {Murphy}, {Nakanishi}, {Pisano}, \&
  {Staveley-Smith}}]{HI4PICollaboration_16}
{HI4PI Collaboration}, {Ben Bekhti}, N., {Fl{\"o}er}, L., {et~al.} 2016, \aap,
  594, A116, \dodoi{10.1051/0004-6361/201629178}

\bibitem[{Hlavacek-Larrondo {et~al.}(2015)Hlavacek-Larrondo, McDonald, Benson,
  Forman, Allen, Bleem, Ashby, Bocquet, Brodwin, Dietrich, Jones, Liu,
  Reichardt, Saliwanchik, Saro, Schrabback, Song, Stalder, Vikhlinin, \&
  Zenteno}]{larondo_15}
Hlavacek-Larrondo, J., McDonald, M., Benson, B.~A., {et~al.} 2015, The
  Astrophysical Journal, 805, 35, \dodoi{10.1088/0004-637x/805/1/35}

\bibitem[{{Hudson} {et~al.}(2010){Hudson}, {Mittal}, {Reiprich}, {Nulsen},
  {Andernach}, \& {Sarazin}}]{hudson_10}
{Hudson}, D.~S., {Mittal}, R., {Reiprich}, T.~H., {et~al.} 2010, \aap, 513,
  A37, \dodoi{10.1051/0004-6361/200912377}

\bibitem[{{Macconi} {et~al.}(2022){Macconi}, {Grandi}, {Gitti}, {Vignali},
  {Torresi}, \& {Brighenti}}]{macconi_22}
{Macconi}, D., {Grandi}, P., {Gitti}, M., {et~al.} 2022, \aap, 660, A32,
  \dodoi{10.1051/0004-6361/202143024}

\bibitem[{McNamara \& Nulsen(2007)}]{mcnamara_07}
McNamara, B., \& Nulsen, P. 2007, Annual Review of Astronomy and Astrophysics,
  45, 117–175, \dodoi{10.1146/annurev.astro.45.051806.110625}

\bibitem[{McNamara \& Nulsen(2012)}]{mcnamara_12}
McNamara, B.~R., \& Nulsen, P. E.~J. 2012, New Journal of Physics, 14, 055023,
  \dodoi{10.1088/1367-2630/14/5/055023}

\bibitem[{Mohr {et~al.}(1999)Mohr, Mathiesen, \& Evrard}]{mohr_99}
Mohr, J.~J., Mathiesen, B., \& Evrard, A.~E. 1999, \apj, 517, 627,
  \dodoi{10.1086/307227}

\bibitem[{Owers {et~al.}(2009)Owers, Nulsen, Couch, \& Markevitch}]{Owers_09}
Owers, M.~S., Nulsen, P. E.~J., Couch, W.~J., \& Markevitch, M. 2009, The
  Astrophysical Journal, 704, 1349, \dodoi{10.1088/0004-637X/704/2/1349}

\bibitem[{{Pasini} {et~al.}(2021){Pasini}, {Gitti}, {Brighenti}, {O'Sullivan},
  {Gastaldello}, {Temi}, \& {Hamer}}]{pasini_21}
{Pasini}, T., {Gitti}, M., {Brighenti}, F., {et~al.} 2021, \apj, 911, 66,
  \dodoi{10.3847/1538-4357/abe85f}

\bibitem[{{Pasini} {et~al.}(2019){Pasini}, {Gitti}, {Brighenti}, {Temi},
  {Amblard}, {Hamer}, {Ettori}, {O'Sullivan}, \& {Gastaldello}}]{pasini_19}
---. 2019, \apj, 885, 111, \dodoi{10.3847/1538-4357/ab4808}

\bibitem[{Peterson \& Fabian(2006)}]{peterson_06}
Peterson, J., \& Fabian, A. 2006, Physics Reports, 427, 1,
  \dodoi{https://doi.org/10.1016/j.physrep.2005.12.007}

\bibitem[{{Pinto} {et~al.}(2014){Pinto}, {Fabian}, {Werner}, {Kosec},
  {Ahoranta}, {de Plaa}, {Kaastra}, {Sanders}, {Zhang}, \&
  {Finoguenov}}]{pinto_14}
{Pinto}, C., {Fabian}, A.~C., {Werner}, N., {et~al.} 2014, \aap, 572, L8,
  \dodoi{10.1051/0004-6361/201425270}

\bibitem[{{Sanders}(2006)}]{sanders_06}
{Sanders}, J.~S. 2006, \mnras, 371, 829,
  \dodoi{10.1111/j.1365-2966.2006.10716.x}

\bibitem[{{Sanderson} {et~al.}(2009){Sanderson}, {Edge}, \&
  {Smith}}]{sanderson_09}
{Sanderson}, A. J.~R., {Edge}, A.~C., \& {Smith}, G.~P. 2009, \mnras, 398,
  1698, \dodoi{10.1111/j.1365-2966.2009.15214.x}

\bibitem[{{Su} {et~al.}(2017){Su}, {Nulsen}, {Kraft}, {Roediger}, {ZuHone},
  {Jones}, {Forman}, {Sheardown}, {Irwin}, \& {Randall}}]{su_17}
{Su}, Y., {Nulsen}, P. E.~J., {Kraft}, R.~P., {et~al.} 2017, \apj, 851, 69,
  \dodoi{10.3847/1538-4357/aa989e}

\bibitem[{{Sutherland} \& {Dopita}(1993)}]{sutherland_93}
{Sutherland}, R.~S., \& {Dopita}, M.~A. 1993, \apjs, 88, 253,
  \dodoi{10.1086/191823}

\bibitem[{{Ubertosi} {et~al.}(2023){Ubertosi}, {Gitti}, \&
  {Brighenti}}]{ubertosi_23}
{Ubertosi}, F., {Gitti}, M., \& {Brighenti}, F. 2023, \aap, 670, A23,
  \dodoi{10.1051/0004-6361/202244023}

\bibitem[{{Ubertosi} {et~al.}(2021{\natexlab{a}}){Ubertosi}, {Gitti},
  {Torresi}, {Brighenti}, \& {Grandi}}]{ubertosi2021a}
{Ubertosi}, F., {Gitti}, M., {Torresi}, E., {Brighenti}, F., \& {Grandi}, P.
  2021{\natexlab{a}}, \mnras, 503, 4627, \dodoi{10.1093/mnras/stab819}

\bibitem[{{Ubertosi} {et~al.}(2021{\natexlab{b}}){Ubertosi}, {Gitti},
  {Brighenti}, {Brunetti}, {McDonald}, {Nulsen}, {McNamara}, {Randall},
  {Forman}, {Donahue}, {Ignesti}, {Gaspari}, {Ettori}, {Feretti}, {Blanton},
  {Jones}, \& {Calzadilla}}]{ubertosi2021b}
{Ubertosi}, F., {Gitti}, M., {Brighenti}, F., {et~al.} 2021{\natexlab{b}}, The
  Astrophysical Journal Letters, 923, L25, \dodoi{10.3847/2041-8213/ac374c}

\bibitem[{Valentini \& Brighenti(2015)}]{valentini_15}
Valentini, M., \& Brighenti, F. 2015, Monthly Notices of the Royal Astronomical
  Society, 448, 1979, \dodoi{10.1093/mnras/stv090}

\bibitem[{{van den Bosch} {et~al.}(2005){van den Bosch}, {Weinmann}, {Yang},
  {Mo}, {Li}, \& {Jing}}]{bosch_05}
{van den Bosch}, F.~C., {Weinmann}, S.~M., {Yang}, X., {et~al.} 2005, \mnras,
  361, 1203, \dodoi{10.1111/j.1365-2966.2005.09260.x}

\bibitem[{{ZuHone} {et~al.}(2016){ZuHone}, {Miller}, {Simionescu}, \&
  {Bautz}}]{zuhone_16b}
{ZuHone}, J.~A., {Miller}, E.~D., {Simionescu}, A., \& {Bautz}, M.~W. 2016,
  \apj, 821, 6, \dodoi{10.3847/0004-637X/821/1/6}

\bibitem[{ZuHone \& Roediger(2016)}]{zuhone_16}
ZuHone, J.~A., \& Roediger, E. 2016, Journal of Plasma Physics, 82,
  \dodoi{10.1017/s0022377816000544}

\bibitem[{{ZuHone} {et~al.}(2019){ZuHone}, {Zavala}, \&
  {Vogelsberger}}]{zuhone_19}
{ZuHone}, J.~A., {Zavala}, J., \& {Vogelsberger}, M. 2019, \apj, 882, 119,
  \dodoi{10.3847/1538-4357/ab321d}

\end{thebibliography}
\bibliographystyle{aasjournal}



\end{document}